\documentclass[prb,twocolumn,showpacs,10pt]{revtex4-1}
\usepackage{textcomp}
\usepackage{amsmath}
\usepackage{amsfonts}
\usepackage{calc}
\usepackage{mathrsfs}
\usepackage{amssymb}
\usepackage{amsmath}
\usepackage{array}
\usepackage{color}
\usepackage{bm}
\usepackage{graphicx}
\usepackage{hyperref}
\newcommand{\floor}[1]{\left\lfloor#1\right\rfloor}

\begin{document}
\title{Effects of electron-phonon coupling on  Landau levels in graphene} 
\author{Adam Pound$^1$} 
\author{J.P. Carbotte$^{2,3}$}
\author{E.J. Nicol$^1$}
\affiliation{$^1$Department of Physics, University of Guelph, Guelph, Ontario, Canada, N1G 2W1}
\affiliation{$^2$Department of Physics and Astronomy, McMaster University, Hamilton, Ontario, Canada, L8S 4M1}
\affiliation{$^3$The Canadian Institute for Advanced Research, Toronto, Ontario, Canada, M5G 1Z8}
\pacs{73.22.Pr, 71.70.Di, 63.22.Rc}
\date{\today}

\begin{abstract}
We calculate the density of states (DOS) in graphene for electrons coupled to a phonon in an external magnetic field. We find that coupling to an Einstein mode of frequency $\omega_E$ not only shifts and broadens the Landau levels (LLs), but radically alters the DOS by introducing a new set of peaks at energies $E_n\pm\omega_E$, where $E_n$ is the energy of the $n$th LL. If one of these new peaks lies sufficiently close to a LL, it causes the LL to split in two; if the system contains an energy gap, a LL may be split in three. The new peaks occur outside the interval $(-\omega_E,\omega_E)$, leaving the LLs in that interval largely unaffected. If the chemical potential is greater than the phonon frequency, the zeroth LL lies outside the interval and can be split, eliminating its association with a single Dirac point. We find that coupling to an extended phonon distribution such as a Lorentzian or Debye spectrum does not qualitatively alter these results.
\end{abstract}
\maketitle

\section{Introduction}
Graphene, a single sheet of carbon atoms arranged in a honeycomb lattice, has aroused much interest in recent years for its many remarkable properties.\cite{Castro-Neto:09, Abergel:10, Geim:07} Most of these properties follow from the fact that at low energies, the charge carriers obey the Dirac equation for massless fermions. Their dispersion relation is linear in momentum, with two bands, valence and conduction, that each form a cone called the Dirac cone, the tips of which meet at a point called the Dirac point. Angle-resolved photoemission spectroscopy (ARPES), which provides a direct measure of the electronic dispersion curves, has verified this band structure.\cite{Bostwick:07,Zhou:08,Bianchi:10} From the dispersion, other interesting characteristics can be calculated: for example, an electronic density of states that increases linearly out of the Dirac point, in striking contrast with the constant value found in conventional metals; and an optical conductivity that displays an intraband Drude peak at zero frequency, followed by a sharp rise to a constant interband contribution at twice the chemical potential $\mu$.\cite{Gusynin:06b, Gusynin:09} These predictions, too, have been confirmed, in scanning tunneling spectroscopy (STS)\cite{Li-G:09,Miller:09} and optical measurements.\cite{Li:08,Orlita:10} However, all of the above experiments have also demonstrated that the simple picture of independent fermions must be modified by many-body renormalizations due to electron-electron\cite{Bostwick:10, Grushin:09, Peres:06, Brar:10,Park:09} and electron-phonon\cite{Brar:10,Stauber:08, Bostwick:07, Zhou:08, Bianchi:10, Li-G:09, Miller:09, Nicol:09, Park:07,Park:09} interactions. In particular, sharp kinks are observed in the linear dispersion\cite{Bostwick:07,Zhou:08,Bianchi:10} and DOS,\cite{Li-G:09} which have been identified as the result of coupling to a phonon. And optical measurements\cite{Li:08} find significant absorption in the region between the Drude peak and the constant background, a phenomenon thought to be at least partially caused by the electron-phonon interaction.\cite{Stauber:08, Carbotte:10, Gusynin:09} Similar absorption absent from the independent-particle prediction\cite{Nicol:08} has also been observed in bilayer systems.\cite{Li-Z:09}

These many-body effects may potentially be amplified, and new ones created, by applying a magnetic field to the system. In a field $B$ perpendicular to the graphene sheet, the linear energy bands are replaced by a set of discrete Landau levels with spacing proportional to $\sqrt{|n|B}$ for integral $n$. The effect of this quantization on the DOS\cite{Sharapov:04} and optical conductivity\cite{Gusynin:07a,Gusynin:07b} have been calculated in an independent-particle picture, and those predictions have been verified in experiment.\cite{Castro-Neto:09, Miller:09, Miller:10, Zeng:10, Song:10, Jiang:07} But we can expect these leading-order predictions to be modified by an assortment of many-body effects made possible by the more complex electronic structure of the quantized system. In particular, we can expect to observe a rich interplay between the LLs and phonons, because for typical magnetic fields used in experiment, the LL spacing is on the scale of the phonon energies in graphene. Our goal here is to extend the independent-particle predictions by taking a first step, within a simple model, toward characterizing that interplay as it manifests in the electronic DOS.

In a companion paper,\cite{Pound:11} we showed that electron-phonon coupling induces a new set of peaks in the DOS, which we called phonon peaks, and that these peaks cause nearby Landau levels to split in two. We also showed that this can explain a disruption of the LLs at a characteristic frequency (which we identified with the phonon frequency) in STS data.\cite{Miller:09} In the present paper, we complement those results by providing analytical calculations and considering a wider range of situations, focusing on descriptions of the effects of coupling rather than on quantitative predictions. We present the necessary formalism in Sec.~\ref{formalism}, where we determine the locations of all phonon peaks by calculating the self-energy in the case of coupling to a single Einstein mode of frequency $\omega_E$. In Sec.~\ref{without chemical potential} we show numerical results for the renormalized DOS in the case of zero chemical potential, emphasizing the locations of the renormalized LLs and their interactions with the phonon peaks. We also present some results in the case of a Lorentzian distribution of phonons, showing that features are smeared but qualitatively unchanged from those of the Einstein model. In Sec.~\ref{with chemical potential} we discuss the case of finite chemical potential, analyzing the locations of the phonon peaks, which become discontinous functions of $B$, and the possibility of splitting the zeroth LL when $\mu$ is greater than $\omega_E$. In Sec.~\ref{other}, we discuss two more scenarios in which electron-phonon coupling induces new structures in the DOS: the presence of an energy gap, and coupling to an acoustic phonon. We conclude in Sec.~\ref{conclusion}.

\section{Formalism}\label{formalism}
\subsection{Governing equations and parameters}
In a magnetic field $B$, the bare electronic density of states with zero chemical potential consists of a line at each Landau level:\cite{Sharapov:04}
\begin{align}
N_{\rm bare}(\omega)&=\frac{1}{2}N_0B\theta(W-|\omega|)\bigg[\delta(\omega+\Delta)+\delta(\omega-\Delta)\nonumber\\
&\quad+2\sum_{n\neq0}\delta(\omega-M_n)\bigg],\label{N_bare}
\end{align}
where $M_n={\rm sgn}(n)\sqrt{2|n|B+\Delta^2}$ is the energy of the $n$th level and $W$ is a high-energy cutoff. The overall scale of the DOS is set by $N_0\equiv\frac{2}{\pi\hbar^2v_F^2}$ and $B$, where the latter is in units of energy squared. To express $B$ in units of teslas, we let $B\to eB\hbar v_F^2$ (in SI units). We have allowed for the possibility of an energy gap $\Delta$ about the Dirac point, which splits the $n=0$ LL. Such a gap is closely related to an asymmetry between the electron densities on the two sublattices.\cite{Khveshchenko:01,Gusynin:06a} Accounting for the cutoff $W$, if $\Delta=0$ then the bare DOS has peaks at $\omega=M_n$ for $-\floor{\frac{W^2}{2B}}\leq n\leq\floor{\frac{W^2}{2B}}$, where $\floor{x}$ is the largest integer smaller than $x$. If a finite gap opens, the zeroth peak is split into two, located at $\omega=\pm\Delta$, while the other peaks are only shifted by the inclusion of $\Delta^2$ in $M_n$.

\begin{figure}[tb]
\begin{center}
\includegraphics[width=\columnwidth]{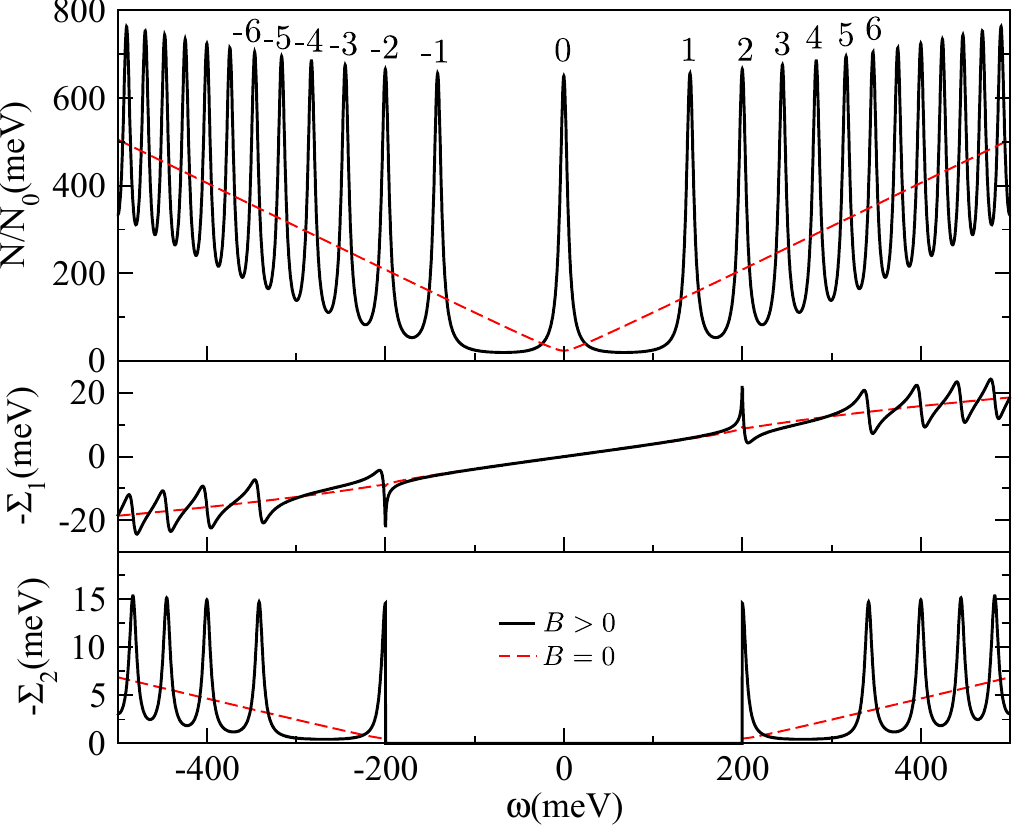}
\end{center}
\caption{(Color online) The bare DOS with a magnetic field (solid black) and without (dashed red), along with the corresponding real and imaginary parts of the self-energy calculated from those bare densities. Integers label the bare Landau levels. Here $A=50$meV, $\omega_E=200$meV, $\Gamma=5$meV, and the black curves are calculated for $B=15.2$T. Note that $N_0$ has dimensions of inverse energy squared, so the quantity $N/N_0$ has dimensions of energy rather than the density of states' usual dimensions of inverse energy. \label{Fig:B_and_noB}}
\end{figure}

In practice, the lines in the DOS are broadened into Lorentzians as $\delta(x)\to\frac{1}{\pi}\frac{\Gamma}{x^2+\Gamma^2}$. For simplicity, we will take $\Gamma$ to be constant; in a more comprehensive calculation, it would be determined by the imaginary part of the full electron self-energy, and it would depend on both $\omega$ and $n$. To orient the reader, we show the typical form of the bare DOS with finite $\Gamma$ in the upper frame of Fig.~\ref{Fig:B_and_noB}. The bare DOS at finite $B$ (shown in solid black) consists of a slowly varying piece, given by the DOS at $B=0$ (shown in dashed red), plus oscillations due to the magnetic field. The slowly varying piece is contributed by the sum of tails of Lorentzians, which will be significant in what follows.

As we shall see, the electron-phonon interaction radically alters this set of peaks. The renormalized DOS is given by the convolution of the bare DOS with a Lorentzian broadened and shifted by the electron-phonon self-energy:
\begin{align}
N(\omega)&=\int^\infty_{-\infty}d\omega'N_{\rm bare}(\omega')\nonumber\\
&\quad\times\frac{1}{\pi}\frac{\Gamma-\Sigma_2(\omega)}{\left[\omega-\Sigma_1(\omega)+\mu-\omega'\right]^2+\left[\Gamma-\Sigma_2(\omega)\right]^2},
\end{align}
where $\Sigma=\Sigma_1+i\Sigma_2$ is the self-energy. One can write this simply as a sum of Lorentzians, 
\begin{align}
N(\omega)&=N_0B\frac{1}{2\pi}\theta(W-|\omega-\Sigma_1(\omega)+\mu|)\nonumber\\
&\quad\times\bigg\lbrace \frac{\Gamma-\Sigma_2(\omega)}{[\omega-\Sigma_1(\omega)+\mu-\Delta]^2+[\Gamma-\Sigma_2(\omega)]^2}\nonumber\\
&\quad+\frac{\Gamma-\Sigma_2(\omega)}{[\omega-\Sigma_1(\omega)+\mu+\Delta]^2+[\Gamma-\Sigma_2(\omega)]^2}\nonumber\\
&\quad+\sum_{n\neq0}\frac{2\left[\Gamma-\Sigma_2(\omega)\right]}{[\omega-\Sigma_1(\omega)+\mu-M_n]^2+[\Gamma-\Sigma_2(\omega)]^2}\bigg\rbrace.\label{DOS_Lorentzians}
\end{align}
Alternatively, following Sharapov et al.,\cite{Sharapov:04} one can evaluate it to find
\begin{align}
N(\omega) &= \frac{N_0}{\pi}\theta\left(W-|\omega-\Sigma_1(\omega)+\mu|\right)\Bigg\lbrace\left[\Gamma-\Sigma_2(\omega)\right]\ln\frac{W^2}{2B}\nonumber\\
&\quad
-{\rm Im}\left[\tilde\omega\psi\left(\frac{\Delta^2-\tilde\omega^2}{2B}\right) +\frac{B\tilde\omega}{\Delta^2-\tilde\omega^2}\right]\Bigg\rbrace,\label{DOS}
\end{align}
where $\tilde\omega\equiv\omega-\Sigma_1(\omega)+\mu+i\Gamma-i\Sigma_2(\omega)$ and $\psi$ is the digamma function. This latter form of $N$ is more convenient for numerical implementation.

At zero temperature, the self-energy is given by\cite{Nicol:09,Carbotte:10}
\begin{align}
\Sigma(\omega) &= \frac{1}{W}\int_0^\infty d\nu\alpha^2F(\nu)\int^\infty_{-\infty}d\omega'\frac{N(\omega')}{N_0}\nonumber\\
&\quad\times\left[\frac{\theta(\omega')}{\omega-\nu-\omega'+i0^+}+\frac{\theta(-\omega')}{\omega+\nu-\omega'+i0^+}\right],\label{Sigma}
\end{align}
where $\alpha^2F$ is the electron-phonon spectral density. Note that, for simplicity, we will not take into account corrections due to the phonon self-energy as
modified by interaction with Dirac electrons, which have been previously calculated theoretically\cite{Ando:06b,Ando:07,Goerbig:07}
and seen in experiment.\cite{Faugeras:09,Yan:10}

Equations \eqref{DOS} and \eqref{Sigma} are solved iteratively, obtaining the $n$th iteration $N^{(n)}$ from $\Sigma^{(n)}$, and $\Sigma^{(n)}$ from $N^{(n-1)}$. We begin this process with the noninteracting DOS, $N^{(0)}$, obtained by letting $\omega\to\omega+\mu_0$ in $N_{\rm bare}$, where $\mu_0$ is the bare chemical potential. This is equivalent to setting $\Sigma=\Sigma^{(0)}\equiv0$ and $\mu=\mu_0$ in Eq.~\eqref{DOS_Lorentzians} or \eqref{DOS}. In subsequent iterations, the noninteracting chemical potential is altered to the interacting one according to $\mu=\mu_0+\Sigma_1(0)$, which preserves the total number of electrons\cite{Luttinger:60} and keeps fixed the value of the DOS at the Fermi energy.\cite{Carbotte:10} Since we wish only to characterize the general behavior of the renormalized Landau levels and the locations of any new peaks, throughout most of this paper we stop after one iteration, which allows us to cleanly identify the effects of renormalization. We will discuss the effects of further iteration in Sec.~\ref{iterations}

Again only to orient the reader, we show the typical form of the first iteration of the self-energy, calculated from the bare DOS, in the lower two frames of Fig.~\ref{Fig:B_and_noB}. As with the DOS, the self-energy consists of a slowly varying piece equivalent to the $B=0$ self-energy, plus oscillations due to the magnetic field. Here and throughout most of this paper, we adopt an Einstein model for the phonon distribution, given by a single phonon at frequency $\omega_E$: $\alpha^2F(\nu)=A\delta(\nu-\omega_E)$, where $A$ defines the strength of the coupling. As shown by Park et al.,\cite{Park:07} if $\omega_E$ is chosen to be $200$meV, the energy of the $E_{2g}$ mode in graphene, then in the absence of a magnetic field this model provides a phonon-induced electron self-energy in excellent agreement with that from a full first-principles calculation. Here we assume that it remains a good approximation in the case of a finite field. Although the modes in the phonon spectrum may be weighted differently in the presence of a magnetic field---due to selection rules in the matrix elements that come from the vertices of the self-energy diagram---the Einstein model has already proven to yield qualitatively useful statements about the effects of electron-phonon coupling on LLs.\cite{Pound:11} More importantly, since the self-energy simply sums over the couplings to all phonons, studying coupling to a single phonon provides a simple means of understanding and characterizing the effects of coupling to any distribution of them, regardless of their weighting.

Although our goal is to analyze qualitative effects of coupling to a phonon, rather than to make quantitative predictions, we will for the most part use experimentally relevant values of our parameters. For the Fermi velocity, we will always use the typical value $v_F=10^6$m/s.\cite{Castro-Neto:09} In the Einstein model, we will almost exclusively use $\omega_E=200$meV, in accord with the results of Park et al.\cite{Park:07} To match to experiment,\cite{Pound:11} we used a coupling strength $A\simeq100$meV, but here we will generally use a larger value of 250meV in order to make qualitative effects more distinct. For the residual scattering parameter $\Gamma$, we will mostly use a value of 5meV, which lies in the middle of the range $\simeq0.02$--10meV derived from transport measurements.\cite{Jiang:07,Tan:07,Bolotin:08,Neugebauer:09} Ando has shown that a value of this order\cite{Ando:06b} is consistent with broadening due to impurity scattering.\cite{Ando:06a} For the high-energy cutoff we use $W=\sqrt{\pi\sqrt 3}t$, where $t$ is the nearest-neighbor hopping parameter; this ensures that the number of states in the Dirac cones equals the number in the first Brillouin zone. Specifically, we use $W=7$eV, corresponding to the typical value $t\simeq 3$eV. However, the value of this cutoff is largely arbitrary. We can see from Eq.~\eqref{DOS} that it has only a marginal impact on the bare DOS, through the addition of a constant term. And its impact on the dressed DOS is governed directly by the magnitude of the self-energy. The self-energy itself is more strongly affected by $W$, because $W$ determines the limits of integration in Eq.~\eqref{Sigma}. Changing the limits of integration not only determines the total number of phonon peaks, but also the magnitude and slope of the slowly varying piece (i.e., the $B=0$ part) of the self-energy. The factor of $1/W$ only partially cancels this effect. However, both this and the overall effect on the dressed DOS can always be precisely compensated for by changing $A$, since $A$ appears as an overall prefactor in the self-energy. (We have verified this numerically.) Therefore, our model is independent of our choice of $W$ in the sense that for a given value of $W$, any self-energy and dressed DOS can be arrived at. One must only keep in mind that the value of $A$ is not meaningful in itself, since it depends on the choice of $W$.

\subsection{Analytical solutions and peak locations}
Before presenting numerical results for the dressed DOS, we provide an an analytical deduction of its peak structure, which will provide the foundation for the remainder of the paper. We begin by noting that the renormalized Landau levels are located at the peaks of the Lorentzians in Eq.~\eqref{DOS_Lorentzians}, where the term $\omega-\Sigma_1(\omega)+\mu-M_n$ in their denominator vanishes. (Equivalently, the peaks lie at energies where the real part of the argument of $\psi$ in Eq.~\eqref{DOS} is a nonpositive integer.) The energies of these levels we denote by $E_n$, which satisfies $E_n-\Sigma_1(E_n)+\mu=M_n$. To determine the locations of any other peaks, we examine the structure of the self-energy. In the Einstein model, the imaginary part of $\Sigma$ can be evaluated immediately to find
\begin{align}
\Sigma_2&=-\frac{\pi A}{WN_0}\left[N(\omega-\omega_E)\theta(\omega-\omega_E)\right.\nonumber\\
&\quad\left.+N(\omega+\omega_E)\theta(-\omega-\omega_E)\right].\label{Sigma2_general}
\end{align}
Substituting $N(\omega)=N^{(0)}(\omega)=N_{\rm bare}(\omega+\mu_0)$ on the right hand side yields the first-order solution $\Sigma^{(1)}_2$. To most easily see the peak structure, we let $\Gamma=0$ for the moment, such that after simplifying the Heaviside functions, $\Sigma^{(1)}_2$ becomes
\begin{align}
\Sigma^{(1)}_2 &=-\frac{\pi AB}{2W}\bigg\lbrace \delta(\omega-\omega_E+\mu_0-\Delta)\theta(-\mu_0+\Delta)\nonumber\\
&\quad +\delta(\omega-\omega_E+\mu_0+\Delta)\theta(-\mu_0-\Delta)\nonumber\\
&\quad +\delta(\omega+\omega_E+\mu_0-\Delta)\theta(\mu_0-\Delta)\nonumber\\
&\quad +\delta(\omega+\omega_E+\mu_0+\Delta)\theta(\mu_0+\Delta)\nonumber\\
&\quad +2\!\!\!\!\sum_{\substack{n\neq0\\|n|\leq n_{\rm max}}}\!\!\!\!\Big[\delta(\omega-\omega_E+\mu_0-M_n)\theta(-\mu_0+M_n)\nonumber\\
&\quad +\delta(\omega+\omega_E+\mu_0-M_n)\theta(\mu_0-M_n)\Big]\bigg\rbrace,\label{Sigma2_sharp}
\end{align}
where $n_{\rm max}=\floor{\frac{W^2}{2B}}$. Next, the real part of $\Sigma^{(1)}$ is given by
\begin{align}
\Sigma^{(1)}_1 &= \frac{A}{W}\mathcal{P}\int^0_{-\infty}\frac{N^{(0)}(\omega')}{N_0}\frac{d\omega'}{\omega+\omega_E-\omega'}\nonumber\\
&\quad +\frac{A}{W}\mathcal{P}\int_0^{\infty}\frac{N^{(0)}(\omega')}{N_0}\frac{d\omega'}{\omega-\omega_E-\omega'}.\label{Sigma1_general}
\end{align}
Again substituting $N^{(0)}(\omega)=N_{\rm bare}(\omega+\mu_0)$ and letting $\Gamma=0$, and taking careful note of when the supports of the delta functions in $N$ lie within the range of integration, we arrive at
\begin{align}
\Sigma^{(1)}_1 &= \frac{AB}{2W}\Bigg\lbrace\frac{\theta(\mu_0+\Delta)}{\omega+\mu_0+\Delta+\omega_E}+\frac{\theta(\mu_0-\Delta)}{\omega+\mu_0-\Delta+\omega_E}\nonumber\\
&\quad +\frac{\theta(-\mu_0-\Delta)}{\omega+\mu_0+\Delta-\omega_E}+\frac{\theta(-\mu_0+\Delta)}{\omega+\mu_0-\Delta-\omega_E} \nonumber\\
&\quad +\sum_{n=-n_{\rm max}}^{n_1-1}\frac{2}{\omega+\mu_0-M_n+\omega_E}\nonumber\\
&\quad +\sum_{n=n_1}^{-1}\frac{2\theta(-\mu_0-\sqrt{2B})}{\omega+\mu_0-M_n-\omega_E} \nonumber\\
&\quad +\sum_{n=1}^{n_2}\frac{2\theta(\mu_0-\sqrt{2B})}{\omega+\mu_0-M_n+\omega_E}\nonumber\\
&\quad +\sum^{n_{\rm max}}_{n=n_2+1}\frac{2}{\omega+\mu_0-M_n-\omega_E}\Bigg\rbrace,\label{Sigma1_sharp}
\end{align}
where $n_1=-\floor{\frac{\mu_0^2}{2B}}\theta(-\mu_0)$ and $n_2=\floor{\frac{\mu_0^2}{2B}}\theta(\mu_0)$.

From these results we find that both $\Sigma_1(\omega)$ and $\Sigma_2(\omega)$ have peaks at $\Delta-\mu_0\pm\omega_E$, $-\Delta-\mu_0\pm\omega_E$, and $M_n-\mu_0\pm\omega_E$---in other words, the energy levels in the bare DOS, but shifted by plus or minus the phonon energy.  We shall refer to these new peaks as phonon peaks and denote their energies as $P_m$, where $m$ is an integer. To distinguish this new set of integers from that labeling the Landau levels, we will add a subscript $p$, as in $m_p$, when labeling energy levels. The values of $P_m$ are summarized in Table~\ref{peaks}. The upper frame shows the position of the $m_p=0$ level, which depends on the relative values of $\mu_0$ and $\Delta$, a phenomenon which will be explored in Sec.~\ref{gap}. The lower frame shows the positions of all other phonon peaks for the two cases $\mu_0<0$ and $\mu_0>0$. In each case the peak locations are divided into two sets, one containing peaks lying to the left of $-\omega_E$, one containing peaks lying to the right of $+\omega_E$---no peaks (including $P_0$) occur inside the interval $(-\omega_E,\omega_E)$. We shall find in what follows that this window, $(-\omega_E,\omega_E)$, plays an important role in determining the general behavior of the DOS: inside the window, the DOS consists simply of shifted LLs; outside of it, the DOS is radically altered by the appearance of the phonon peaks. In addition, the presence of the window affects the dependence of the peaks' position on the three parameters $\mu_0$, $\Delta$, and $\sqrt{2B}$: because no peaks can occur within the window, continuously varying one of these three parameters so as to make a peak pass through the window's boundary at $\omega_E$ or $-\omega_E$ forces the peak to jump discontinuously to the other side of the window. We will explore this phenomenon in detail in Sec.~\ref{jumping}.

\begingroup
\squeezetable
\begin{table}
\caption{Locations of peaks in $\Sigma$ and $N$ induced by the electron-phonon interaction, in addition to the logarithmic peaks at $\omega=\pm\omega_E$. Each of these peaks lies at the position of a bare Landau level displaced by plus or minus the phonon energy. Top: the new peaks generated from the $n=0$ LL. Bottom: the new peaks generated from the $n\neq0$ LLs.}
\begin{ruledtabular}
\begin{tabular}{c}
$\begin{array}{rl}\mu_0<-\Delta: & P_{0^\pm}=\omega_E\pm\Delta-\mu_0\\
-\Delta<\mu_0<\Delta: & P_{0^\pm}=\pm(\omega_E+\Delta)-\mu_0\\
\mu_0>\Delta: & P_{0^\pm}=-\omega_E\pm\Delta-\mu_0\end{array}$\\
\hline\\[-8pt]
$\begin{array}{rl}\mu_0<0: &\!\! \begin{cases} P_{m}=M_m-\mu_0-\omega_E, & m=-\floor{\frac{W^2}{2B}},\ldots,-\floor{\frac{\mu_0^2}{2B}}-1\\
P_{m}=M_m-\mu_0+\omega_E, & m=-\floor{\frac{\mu_0^2}{2B}},\ldots,\floor{\frac{W^2}{2B}},m\neq0\end{cases}\\
\mu_0>0: &\!\! \begin{cases}P_{m}=M_m-\mu_0-\omega_E, & m=-\floor{\frac{W^2}{2B}},\ldots,\floor{\frac{\mu_0^2}{2B}},m\neq0\\
P_{m}=M_m-\mu_0+\omega_E, & m=\floor{\frac{\mu_0^2}{2B}}+1,\ldots,\floor{\frac{W^2}{2B}}\end{cases}\end{array}$
\end{tabular}
\end{ruledtabular}
\label{peaks}
\end{table}
\endgroup

Note that we cannot actually find $N^{(1)}(\omega)$ from the self-energy just derived, because it would be a nonlinear functional of $\delta$ functions. To self-consistently iterate the DOS and self-energy, we must begin with broadened $\delta$ functions in $N_{\rm bare}(\omega)$. For $\Sigma^{(1)}_2$, this has the effect of simply broadening the $\delta$ functions in Eq.~\eqref{Sigma2_sharp} into Lorentzians. For $\Sigma^{(1)}_1$, we must reevaluate Eq.~\eqref{Sigma1_general}. Doing so yields
\begin{align}
\Sigma^{(1)}_1(\omega)&=-\frac{AB}{2\pi W}\Big\lbrace H_\Delta(\omega)+H_{-\Delta}(\omega) +2\!\!\!\!\!\sum_{\substack{n\neq0\\|n|\leq n_{\rm max}}}\!\!\!\!H_{M_n}(\omega)\Big\rbrace,\label{Sigma1_broad}
\end{align}
where
\begin{align}
H_p(\omega) &=-\frac{(\omega+\omega_E+\mu_0-p)\arccos\frac{p-\mu_0}{\sqrt{(p-\mu_0)^2+\Gamma^2}}}{(\omega+\omega_E+\mu_0-p)^2+\Gamma^2}\nonumber\\
&\quad-\frac{(\omega-\omega_E+\mu_0-p)\arccos\frac{\mu_0-p}{\sqrt{(\mu_0-p)^2+\Gamma^2}}}{(\omega-\omega_E+\mu_0-p)^2+\Gamma^2}\nonumber\\
&\quad +\frac{\Gamma\ln\frac{|\omega+\omega_E|}{\sqrt{(p-\mu_0)^2+\Gamma^2}}}{(\omega+\omega_E+\mu_0-p)^2+\Gamma^2}\nonumber\\
&\quad -\frac{\Gamma\ln\frac{|\omega-\omega_E|}{\sqrt{(p-\mu_0)^2+\Gamma^2}}}{(\omega-\omega_E+\mu_0-p)^2+\Gamma^2}.\label{H}
\end{align}
Each Lorentzian in $N^{(0)}(\omega)$ has contributed a function, $H_p(\omega)$, that depends on $\omega+\omega_E$ and $\omega-\omega_E$, as we expect. In the limit $\Gamma\to0$, the $\arccos$ terms in Eq.~\eqref{Sigma1_broad} reduce to Eq.~\eqref{Sigma1_sharp}, recovering the peak structure just described. However, for finite $\Gamma$, $\Sigma^{(1)}_1$ contains logarithmic divergences at $\omega=\pm\omega_E$, which are not present in Eq.~\eqref{Sigma1_sharp}. We can understand this by noting that the logarithmic divergences are present in the $B=0$ case,\cite{Nicol:09,Carbotte:10} and hence must come from the slowly increasing piece of the DOS, rather than from the peaks at the Landau levels. And since the slowly increasing piece arises from the sum of the tails of the Lorentzians, the logarithmic divergence exists only for finite $\Gamma$. Similarly, since $N(\omega)$ nowhere vanishes for finite $\Gamma$, $\Sigma_2$ always sharply changes from zero to the finite value $-\frac{\pi A}{WN_0}N(0)$ at $\omega=\pm\omega_E$, as we can see from Eq.~\eqref{Sigma2_general}.

We are now in a position to understand the general form of $\Sigma$, as shown in the lower two frames of Fig.~\ref{Fig:B_and_noB} for the simplest case of $\mu=\Delta=0$. As discussed above, the self-energy for finite $B$ (solid black curves) consists of a slowly varying piece (dashed red curves), constructed from the $B=0$ piece of the DOS (shown in the upper frame), plus oscillations, constructed from the oscillations in the DOS. In fact, as we see in both the figure and Eq.~\eqref{Sigma2_general}, the oscillations in $\Sigma_2$ are precise images of those in the bare DOS, simply rescaled and translated by the phonon energy. Note the distinctive character of the peaks in $\Sigma_1$ and $\Sigma_2$ at $\pm\omega_E$. At those points, $\Sigma_2$ becomes finite and picks up half an image of the $n=0$ bare Landau level, while in $\Sigma_1$ the logarithmic singularity contributes to the $P_{0^+}$ peak picked up from the $n=0$ bare LL. Since no oscillations occur in the window $(-\omega_E,\omega_E)$, in that region the self-energy in the $B>0$ case cannot be distinguished from that in the $B=0$ case.

Since the DOS has the form of oscillations atop a slowly varying function, each peak in the self-energy will introduce a peak in $N^{(1)}$ via that slowly varying piece. Therefore, from the above analysis we can conclude that the renormalized DOS contains the following peaks: renormalized Landau levels at energies $E_n$ labeled by integers $n$, which are solutions to $E_n-\Sigma_1(E_n)+\mu = M_n$; phonon peaks at energies $P_m$ labeled with integers $m_p$, which arise as images of the LLs in the bare DOS and always lie outside the interval $(-\omega_E,\omega_E)$; and phonon peaks at $\pm\omega_E$, which arise from a logarithmic divergence in $\Sigma_1$ due to the abrupt opening of a new scattering channel at the phonon frequency. Depending on the values of $\mu_0$ and $\Delta$, the zeroth phonon peak $P_0$ may or may not be split in two and may or may not coincide with the logarithmic divergences. Furthermore, as we shall see in Sec.~\ref{splitting}, each of the renormalized LLs may be split by the presence of the phonon peaks.

\section{Density of states for vanishing chemical potential}\label{without chemical potential}

\subsection{Appearance of phonon peaks}

\begin{figure}[tb]
\begin{center}
\includegraphics[width=\columnwidth]{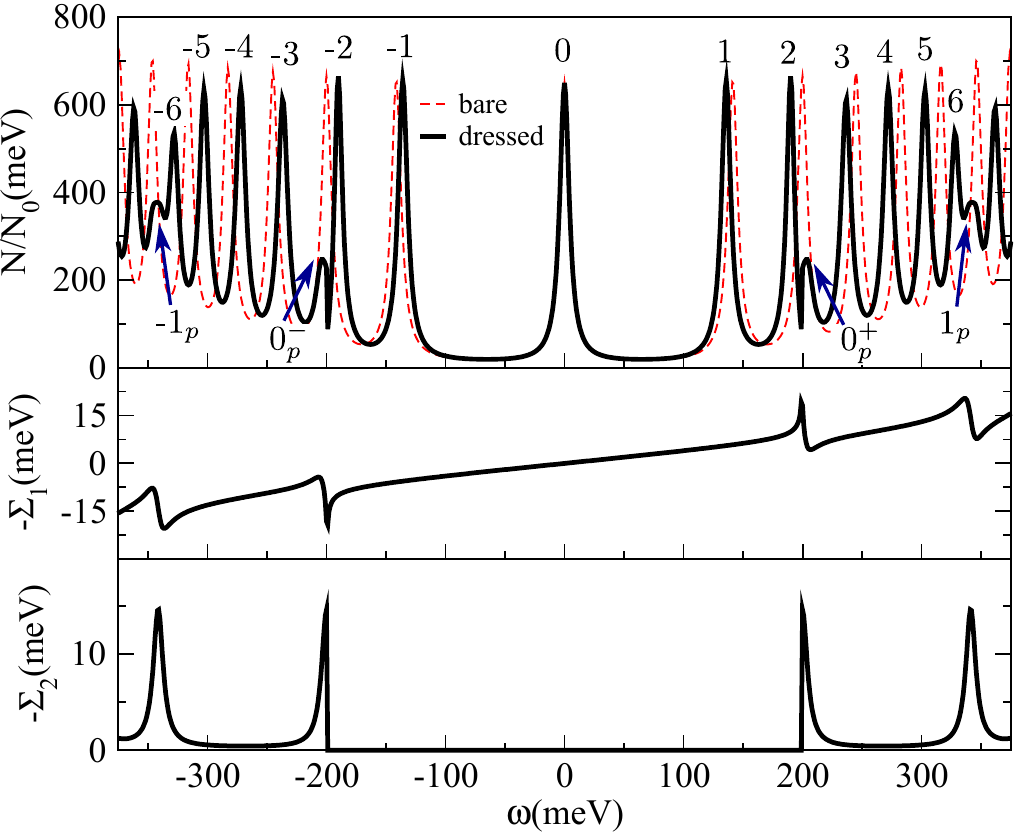}
\caption{(Color online) Density of states and self-energy as a function of energy, for $A=50$meV, $\omega_E=200$meV, $B=15.2$T, and $\Gamma=5$meV. The dashed red curve shows the bare DOS, while the solid black curve shows the DOS renormalized by $\Sigma$. Integers $n$ label the dressed Landau levels. Integers $m_p$ label the phonon peaks, pointed to by blue arrows.\label{Fig:lowA}}
\end{center}
\end{figure}

We begin our numerical analysis of the dressed DOS by examining the appearance of phonon peaks in the simplest case of vanishing $\mu$ and $\Delta$. As discussed above, just as $\Sigma$ picks up the oscillations in the bare DOS, the renormalized DOS picks up those in $\Sigma$. Figure~\ref{Fig:lowA} shows the general effects of this renormalization. Here we use a fairly weak coupling of $A=50$meV, which allows us to easily compare the renormalized DOS (solid black curve in the top frame) to the bare DOS (dashed red curve). We also show the self-energy, in the lower two frames, to make transparent its manifestation in the dressed DOS. For small $\omega$, $-\Sigma_1(\omega)=\omega\lambda^{\rm eff}+O(\omega^2)$, where $\lambda^{\rm eff}\equiv-\frac{\partial\Sigma_1}{\partial\omega}(0)$ is the effective mass renormalization parameter. As seen in the middle frame of the figure, this linear behavior remains a good approximation in the window $(-\omega_E,\omega_E)$. And since $\Sigma_2$ vanishes in that window, the only effect of renormalization therein is to draw the Landau levels closer together, due to the linear change in $\Sigma_1$. For the parameters used here, $\lambda^{\rm eff}=0.04$ and the $n=\pm1$ LLs, located at solutions to $M_{\pm1}=E_{\pm1}-\Sigma_1(E_{\pm1})\simeq E_{\pm1}+\lambda^{\rm eff}E_{\pm1}$, are appropriately shifted down by a factor of $1+\lambda^{\rm eff}$. This shift from $M_{\pm1}$ to $E_{\pm1}=M_{\pm1}/(1+\lambda^{\rm eff})$ can be interpreted as a renormalization of the Fermi velocity in $M_{\pm1}$. Outside the window, above the phonon energy, the levels are still drawn closer together but also broadened, as $\Sigma_2$ becomes finite with the opening of a new scattering channel. More notably, new peaks arise in the DOS, indicated by blue arrows and labeled with integers $m_p$. As described in the previous section, these peaks lie at $M_n\pm\omega_E$, the same locations as the peaks in $\Sigma$. Specifically, $0_p^+$ and $1_p$ lie at $\omega_E$ and $M_1+\omega_E$, respectively, and $0^-_p$ and $-1_p$ lie at $-\omega_E$ and $M_{-1}-\omega_E$.

\begin{figure}[tb]
\begin{center}
\includegraphics[width=\columnwidth]{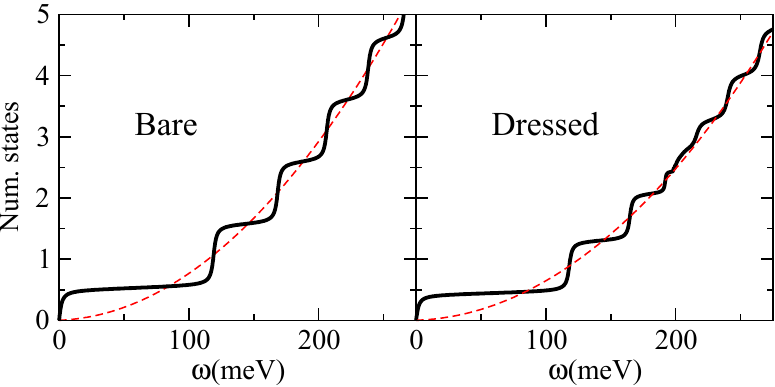}
\end{center}
\caption{(Color online) The sum of states $\int^\omega_0N(\omega')d\omega'$ as a function of energy for $B=15.2$T (solid black curves) and $B=0$ (dashed red curves) in both the bare (left) and dressed (right) cases. The remaining parameters are $A=250$meV, $\omega_E=200$meV, and $\Gamma=5$meV. The number of states is written in units of $10^5$meV$^2N_0$. \label{Fig:num_states}}
\end{figure}

We can also examine how these features in the DOS manifest in the sum of states between $0$ and $\omega$, $\int^\omega_0N(\omega')d\omega'$, as shown in Fig.~\ref{Fig:num_states}. On the left, we see that just as in the DOS, the magnetic field causes the bare $B>0$ sum (solid black curve) to oscillate about the bare $B=0$ sum (dashed red curve). On the right, we see the same general behavior in the dressed sums, except for a disruption around the phonon frequency. Hence, in both cases, adding a magnetic field does not alter the total number of states. 

\subsection{Locations and splitting of dressed Landau levels}\label{splitting}
\begin{figure}[tb]
\begin{center}
\includegraphics[width=\columnwidth]{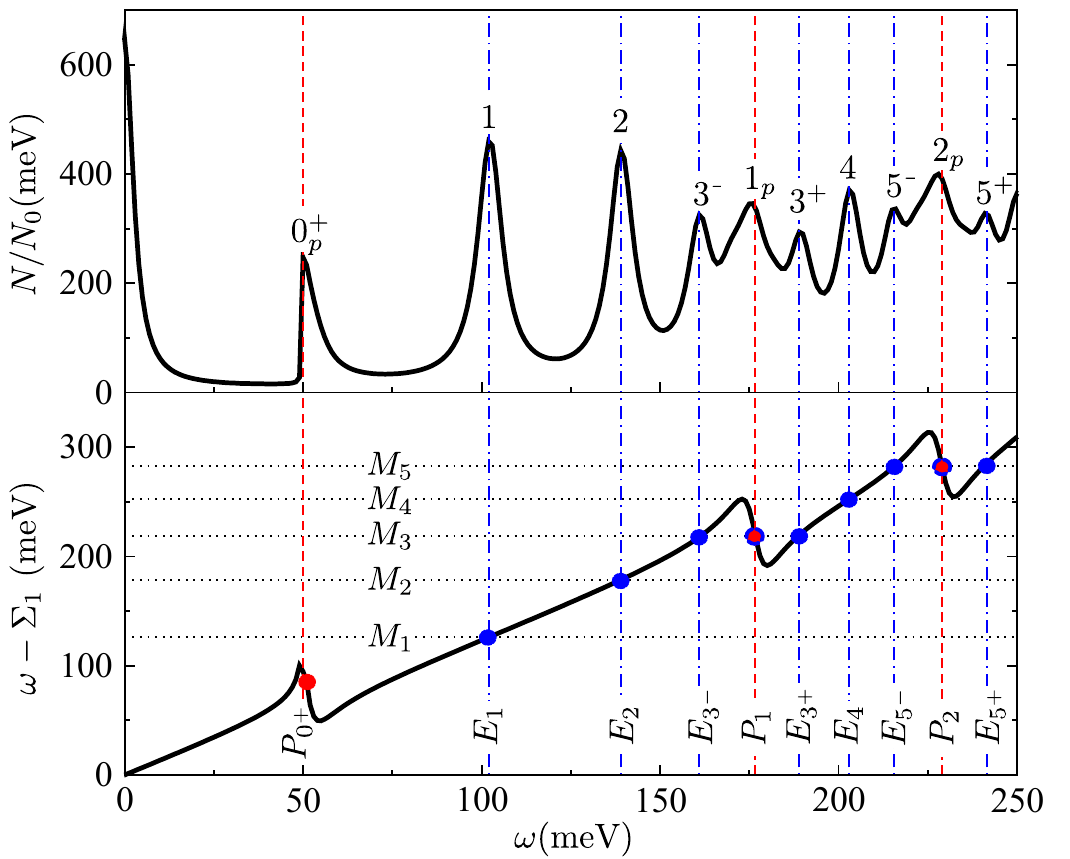}
\end{center}
\caption{(Color online) Locations of peaks in the density of states, as determined by $\omega-\Sigma_1(\omega)$, for $A=250$meV, $\omega_E=50$meV, $B=12.2$T, and $\Gamma=4$meV. Straight lines indicate the bare Landau-level energies $M_n$ (dotted black lines), the energies of renormalized Landau levels $E_n$ (dot-dashed blue), and the positions $P_m$ of phonon peaks (dashed red). Blue circles mark the intersection of $\omega-\Sigma_1$ with $M_n$, which determines the energies $E_n$; since these intersections occur twice for $M_3$ and $M_5$, the third and fifth LLs are split. Red circles mark the center-points of oscillations in $\Sigma$, which determine the positions of phonon peaks. Those red circles outlined in blue also satisfy the equations $\omega-\Sigma_1=M_n$ (for $n=3$ and $n=5$).\label{Fig:low_wE}}
\end{figure}

In Fig.~\ref{Fig:low_wE} we show more precisely how and where the two sets of peaks in the DOS appear, now choosing a stronger, more realistic coupling. We also choose a small phonon frequency, such that we can see the first phonon peak clearly, with no nearby Landau levels. The upper frame shows the DOS and the lower shows a plot of $\omega-\Sigma_1(\omega)$, both in solid black. The energies of the first three phonon peaks, labeled $0_p^+$, $1_p$, and $2_p$ in the upper frame, are indicated by vertical dashed red lines. We see in the bottom frame that these energies correspond to the center point of an oscillation in $\omega-\Sigma_1(\omega)$. As we can see, the first peak is highly asymmetric, since no phonon peaks can occur below $\omega_E$. It is the image of half of the $n=0$ level from the bare DOS; the other half would lie at $-\omega_E$.

However, the plot of $\omega-\Sigma_1(\omega)$ is primarily useful in determining the locations of the Landau levels, rather than the already-known locations of the phonon peaks. To this end, in the lower frame we also plot the bare energies $M_n$, shown with horizontal dotted lines; since the renormalized LLs lie at the solutions to $E_n-\Sigma_1(E_n)=M_n$, their renormalized energies $E_n$ occur at the intersections of these dotted lines with the solid black curve. The energies $E_n$ are then indicated with dashed blue vertical lines passing through the intersection points. Looking at the top frame, we see that those intersection points correctly determine the locations of peaks in the dressed DOS. We also see that two of the LLs, $n=3$ and $n=5$, are split. This occurs due to oscillations in $\Sigma_1$, which allow multiple intersections to occur for a given $M_n$. Because a phonon peak always occurs at the center of an oscillation, the LLs will always be split such that one peak occurs to the left of a phonon peak and one to the right.

This splitting occurs only if a Landau level is sufficiently close to a phonon peak. More precisely, it occurs only when the bare level at $M_n$ lies within the amplitude of an oscillation in $\omega-\Sigma_1(\omega)+\mu$ (including the possibility of finite $\mu$ for generality). Denoting the amplitude of the $m$th oscillation, occurring about $P_m-\Sigma_1(P_m)+\mu$, by $\delta\Sigma_1(P_m)$, we can write the condition for the $n$th LL to be split about the $m$th phonon peak as $\left|M_n-\left[P_m-\Sigma_1(P_m)+\mu\right]\right|<\delta\Sigma_1(P_m)$. We can straightforwardly derive a formula for $\delta\Sigma_1$ from Eqs.~\eqref{Sigma1_broad} and \eqref{H}, as follows. Near the energy $P_m$, we can treat all but the $m$th peak as a smooth background and look at the function $\frac{AB}{\pi W}\frac{\omega-P_m}{(\omega-P_m)^2+\Gamma^2}$ alone. This vanishes at $\omega=P_m$ and has extrema at $\omega=P_m\pm\Gamma$. For sufficiently small $\Gamma$, we can set the $\arccos$ factor that multiplies this equal to $\pi$, so the oscillations have magnitude $\delta\Sigma_1=\frac{AB}{2W\Gamma}$, independent of $m$. Noting that the only effect of $B$ is to create oscillations about the $B=0$ case, we have $\Sigma_1(P_m)=\Sigma_1^{B=0}(P_m)$. After similarly noting $\mu=\mu_0+\Sigma_1^{B=0}(0)$, we can conclude that in general, the $n$th LL will be split in two about $P_m$ when $M_n$ satisfies
\begin{equation}
\big|M_n-\left[P_m-\Delta\Sigma_1(P_m)+\mu_0\right]\big|\lesssim\delta\Sigma_1,\label{splitting_condition}
\end{equation}
where $\Delta\Sigma_1(P_m)\equiv\Sigma_1^{B=0}(P_m)-\Sigma_1^{B=0}(0)$ is the change in $\Sigma_1$ away from its value at the Fermi energy, and $\delta\Sigma_1=\frac{AB}{2W\Gamma}$ is the magnitude of the oscillation in $\Sigma_1$ about its value at $P_m$. With this condition, one can determine whether a level will be split without any numerical computation, since all quantities are known analytically in terms of elementary functions. [The equation for $\Sigma^{B=0}_1(\omega)$ can be found in Carbotte et al.\cite{Carbotte:10}] In addition, the bare chemical potential $\mu_0$ need not be known, since its appearance in $P_m$ cancels its explicit appearance in Eq.~\eqref{splitting_condition}.

\begin{figure}[tb]
\begin{center}
\includegraphics[scale=0.8]{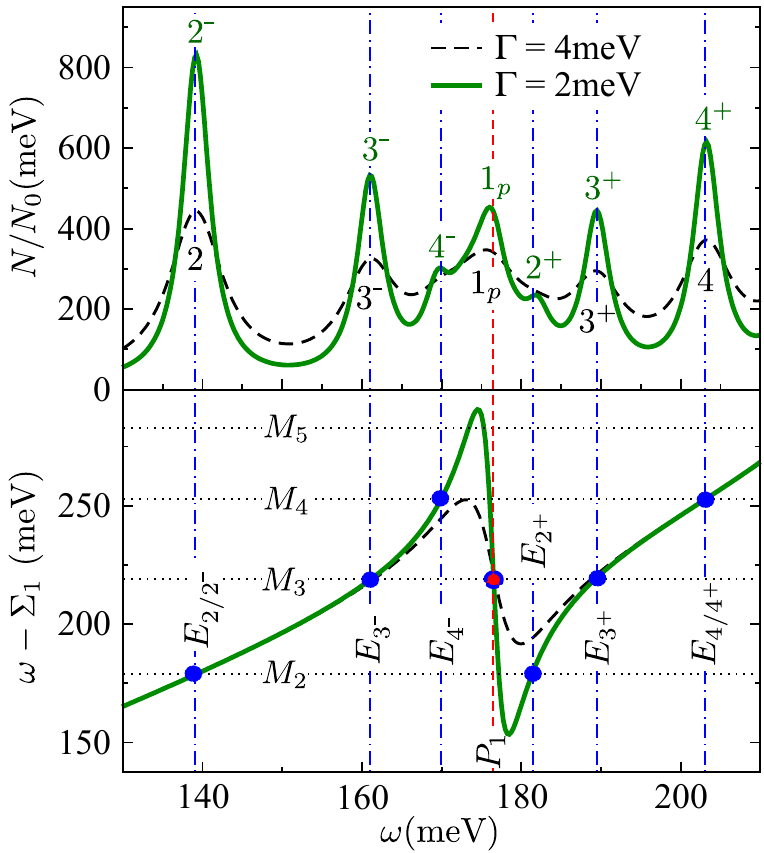}
\end{center}
\caption{(Color online) The effect of $\Gamma$ on the number of peaks in the density of states, showing a small portion of Fig.~\ref{Fig:low_wE}, where the dashed black curves here are identical to the solid black curves therein, and the solid green curves are for a halved value of $\Gamma$. The curve with $\Gamma=2$meV intersects two more $M_n$'s than does the curve with $\Gamma=4$meV, creating two more split Landau levels.\label{Fig:Gamma_splitting}}
\end{figure}

We illustrate the effect of changing $\Gamma$ in Fig.~\ref{Fig:Gamma_splitting}, which shows a small portion of Fig.~\ref{Fig:low_wE} but now with two values of $\Gamma$. The dashed black curves are identical to the curves in the previous figure, while the solid green curves are for a halved value of $\Gamma$. Halving $\Gamma$ doubles $\delta\Sigma_1$, which results in more Landau levels being split. Here the black curve, with $\Gamma=4$meV, intersects only the $n=3$ bare LL energy in the lower frame, so only the $n=3$ level is split in the DOS in the upper frame. The green curve intersects the bare energies $M_2$ through $M_5$, leading to four split levels in the DOS. Note that the intersection with $M_5$ is glancing, such that the $n=5^-$ level is not apparent, instead only influencing the asymmetry of adjacent peaks. Similarly, the intersections on the side of the oscillation nearest $P_m$ (e.g., the unmarked intersections of $M_2$ and $M_4$ with the green curve in the lower frame) do not result in a new peak, because those peaks disappear under the phonon peak.

One should note that this analysis does not apply to the logarithmic peaks at $\pm\omega_E$. Referring to Eq.~\eqref{H}, we see that except in the case when a peak at $P_m$ coincides with a logarithmic peak, the logarithmic peak vanishes for $\Gamma\to0$ rather than diverging as the other peaks do.

\begin{figure}[tb]
\begin{center}
\includegraphics[width=\columnwidth]{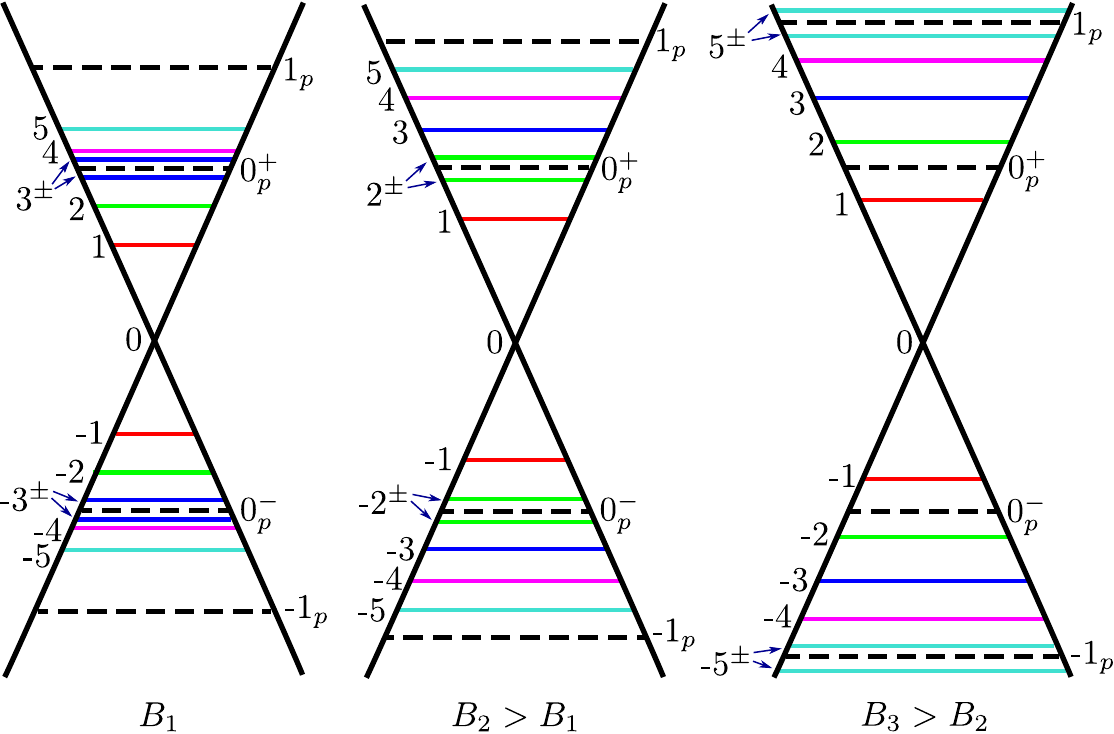}
\end{center}
\caption{(Color online) A schematic depiction of the locations of energy levels on the Dirac cone for three increasing values of $B$. Solid colored lines labeled with integers $n$ or $n^\pm$ indicate Landau levels. Dashed black lines labeled with integers $m_p$ or $m_p^\pm$ indicate energies of phonon peaks. \label{Fig:cones}}
\end{figure}

\begin{figure}[tb]
\begin{center}
\includegraphics[width=\columnwidth]{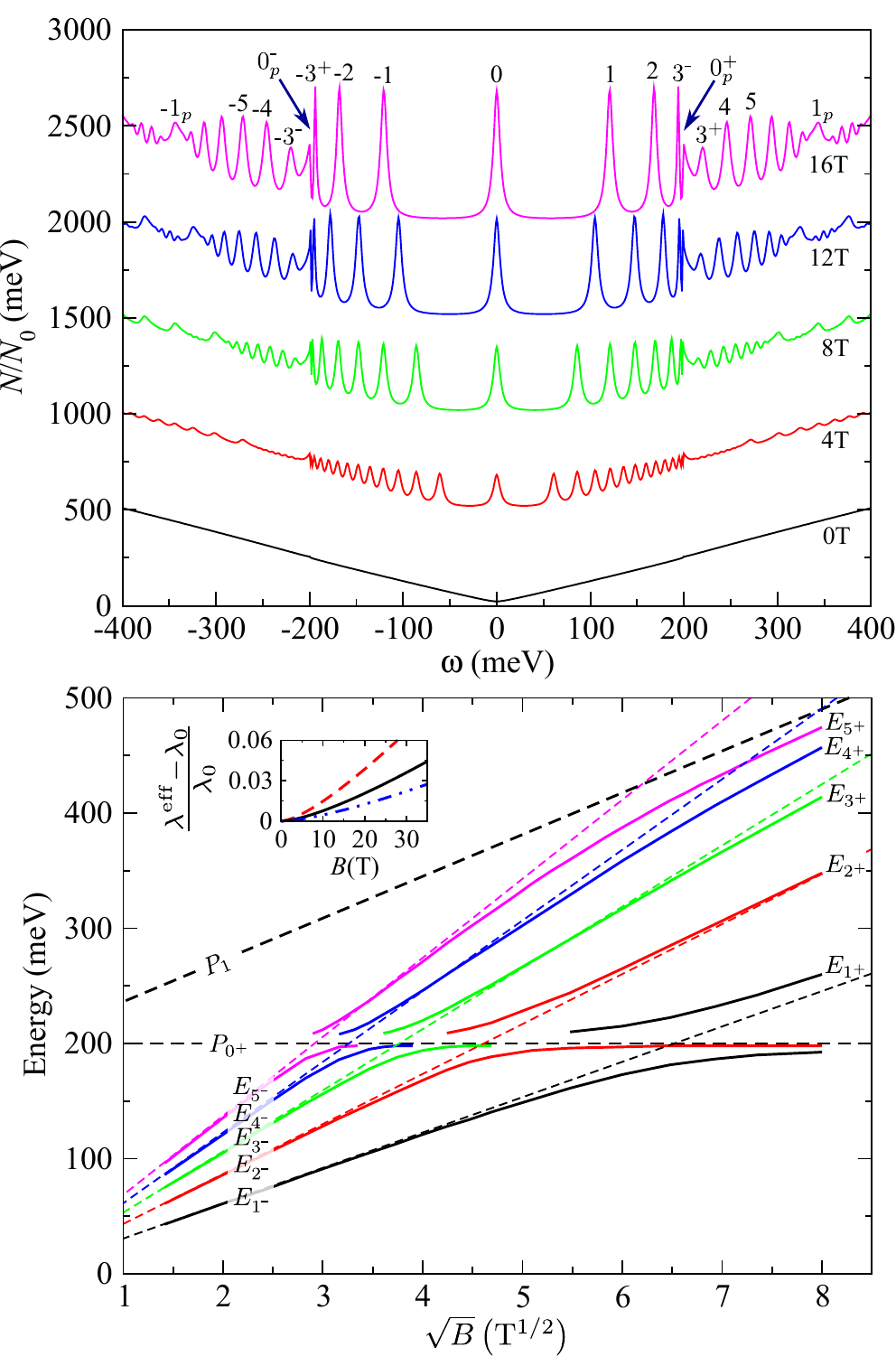}
\end{center}
\caption{(Color online) Top: Density of states as a function of energy for a sequence of magnetic field strengths, with curves offset by 500meV for clarity. Black arrows point to the first two sets of phonon peaks, at $\pm\omega_E$ and $\pm(\omega_E+M_1)$. Integers label the first five sets of Landau levels.  In all cases, $\omega_E=200$meV, $A=250$meV, and $\Gamma=5$meV. Bottom: Locations of peaks in the DOS as a function of $\sqrt{B}$, with the same parameters. The solid curves labeled $E_{n^\pm}$ show the positions of the renormalized Landau levels. The thin dashed lines show $M_n/(1+\lambda_0)$, where $\lambda_0\equiv\lambda^{\rm eff}(B=0)$. The thick black dashed lines show the first two phonon peaks. Inset: the fractional change in $\lambda^{\rm eff}$ away from $\lambda_0$ as a function of $B$, for $\omega_E=150$meV (dashed red curve), $\omega_E=200$meV (solid black), and $\omega_E=250$meV (double-dot dashed blue).\label{Fig:peaks}}
\end{figure}

The foregoing results have all been for a fixed magnetic field. We now consider the effect of varying that field. The general trend is shown schematically in Fig.~\ref{Fig:cones}, which displays the position of the peaks on the Dirac cone for three increasing values of $B$.  As the magnetic field is increased, the dressed Landau levels (shown as colored lines labeled with integers $n$ or $n^\pm$) are pushed further away from the Dirac point. The phonon peaks $\pm m_p$ (shown as dashed black lines) move outward at the same rate as the corresponding LLs, $\pm m$, of which they are images. So in particular, the phonon peaks $0_p^\pm$ remain stationary, since they are images of the $n=0$ LL, which is pinned to the Dirac point. Because the phonon peaks are shifted outward by $\omega_E$ relative to the bare LLs, the LLs pass through them as the magnetic field increases, causing the levels to split before reconverging at higher $B$. In the schematic, we see the $\pm3$ levels split about the $0_p^\pm$ peaks in the first frame, then reconverge in the second; the $\pm2$ levels split about the $0^\pm_p$ peaks in the second, then reconverge in the third; and the $\pm5$ levels split about the $\pm1_p$ peaks in the third.

Figure~\ref{Fig:peaks} displays these effects more precisely. In the upper frame we show a sequence of DOS curves for a range of magnetic fields. We see both the Landau levels and phonon peaks move outward as $B$ increases, with the exception of the zeroth level at the Dirac point and the $0_p^\pm$ phonon peaks at plus or minus the phonon frequency. As the LLs approach the $0_p^\pm$ peaks, they are visibly disrupted, splitting into two peaks of unequal width and then merging back into single peaks once they are sufficiently far out. In the lower frame we show (with solid curves) the energies of the first five (positive $n$) LLs as a function of $\sqrt B$. For comparison, we also show (with dashed lines colored to correspond to the solid ones) the straight lines $M_n/(1+\lambda_0)$, where $\lambda_0\equiv\lambda^{\rm eff}(B=0)$. When sufficiently far from any phonon peaks, the LLs closely follow the straight lines.  Near the phonon peaks, they deviate from this behavior and split in two, as expected. Note that we could calculate curves $M_n/(1+\lambda^{\rm eff})$, where the renormalization factor $1/(1+\lambda^{\rm eff})$ depends on $B$. This would more closely track the true renormalized curves. To determine the magnitude of the correction, in the inset we plot $\lambda^{\rm eff}$ for three values of $\omega_E$. As we would expect, bringing the phonon closer to the Fermi energy increases its effect on the slope of $\Sigma_1$ there, increasing the change in $\lambda^{\rm eff}$ with $B$. However, for the value $\omega_E=200$meV used in the remainder of this paper, all changes are less than 5\% for a very large range of field strengths. So, except near a phonon peak, $M_n/(1+\lambda_0)$ is an accurate approximation for the energy of the $n$th renormalized LL. This result, in combination with the energies of the phonon peaks given in Table~\ref{peaks} and the splitting condition in Eq.~\eqref{splitting_condition}, allows us to calculate the approximate positions of all peaks in the DOS in terms of elementary functions.

\subsection{Broadened phonon spectrum}

\begin{figure}[tb]
\begin{center}
\includegraphics[width=\columnwidth]{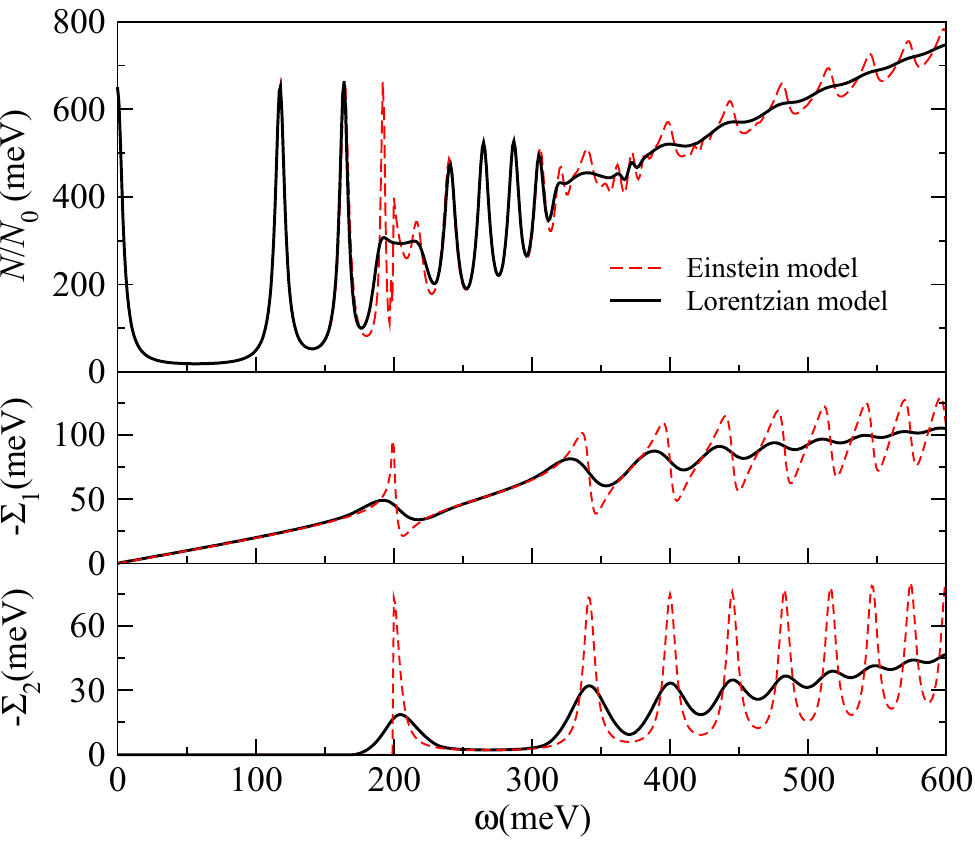}
\end{center}
\caption{(Color online) Density of states and self-energy in the Einstein model (dashed red curves) and in the truncated Lorentzian model (solid black curves) with a magnetic field $B=15.2$T. The Landau levels that are split about $\omega_E$ in the Einstein model are flattened by the broadening of the phonon in the Lorentzian model. In the Einstein model we use $A=250$meV and $\omega_E=200$meV; in the Lorentzian, $\omega_0=200$meV, $A'=555$meV, $\delta=15$meV, and $\delta_c=30$meV. The value of $A'$ is chosen to ensure that $\lambda^{\rm eff}=0.19$ in both models. In both cases, $\Gamma=5$meV.\label{Fig:Lorentz}}
\end{figure}

Thus far, our results have been restricted to the simple Einstein model. In a real physical system, the phonon distribution will be broadened, and this broadening can significantly affect the features of the DOS around the phonon energy. As an example, we consider the truncated Lorentzian distribution of Dogan et al.,\cite{Dogan:03, Dogan:07} given by
\begin{equation}
\alpha^2F(\nu)=\frac{A'}{\pi}\left[\frac{\delta}{(\nu-\omega_0)^2+\delta^2}-\frac{\delta}{\delta_c^2+\delta^2}\right]\theta(\delta_c-|\nu-\omega_0|).
\end{equation}
This is a Lorentzian of width $\delta$ centered at $\omega_0$ and truncated at $\omega_0\pm\delta_c$. In Fig.~\ref{Fig:Lorentz} we present results for this model together with the Einstein model. To compare the results, we set $\omega_0=\omega_E$ and choose $A'$ such that $\lambda^{\rm eff}$ is identical for both models.   We see that in the Lorentzian model, the magnitude of peaks in the self-energy are reduced from that of the Einstein model. More significantly, the peak at $\omega_E$ is smoothed, losing the distinctive one-sided character it has in the Einstein model. This smoothing alters the peak structure around $\omega_E$ in the DOS, merging the split Landau level with the phonon peak to create a flattened structure. As discussed in our companion paper,\cite{Pound:11} this flattened structure can be seen in experimental data.

\subsection{Multiple iterations}\label{iterations}

\begin{figure}[tb]
\begin{center}
\includegraphics[width=\columnwidth]{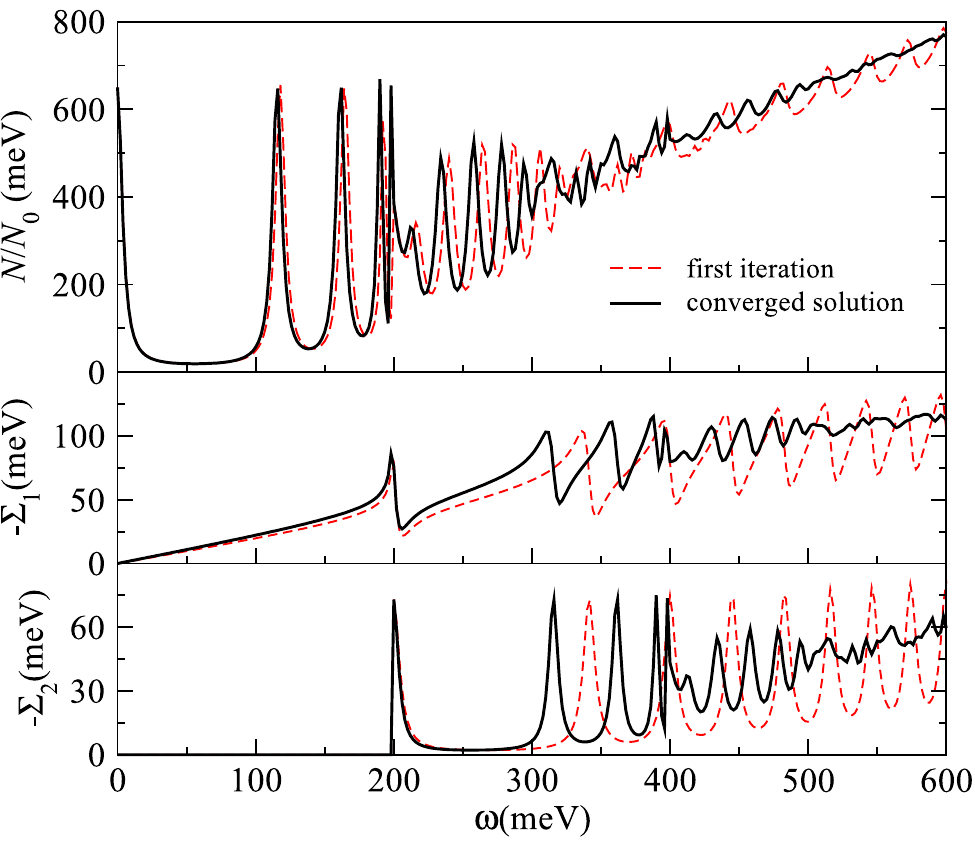}
\end{center}
\caption{(Color online) Density of states and self-energy after one iteration (dashed red curves) and after iterating to convergence (solid black curves), with parameters $B=15.2$T, $A=250$meV, $\omega_E=200$meV, and $\Gamma=5$meV.\label{Fig:iterated}}
\end{figure}

In addition to restricting ourselves to an Einstein model, we have thus far restricted ourselves to one iteration of our self-consistent system of equations. In Fig.~\ref{Fig:iterated} we present the DOS and self-energy after iterating to convergence. The slightly more complicated set of peaks that appears in the DOS can still be divided into Landau levels, which solve $E_n-\Sigma_1(E_n)+\mu=M_n$, and phonon peaks, at which $\Sigma_2$ has its maxima. We see that the converged function $\Sigma_2(\omega)$ contains a perfect image of the dressed, rather than the bare, DOS, such that phonon peaks now lie at $E_n+\omega_E$ rather than $M_n+\omega_E$. This result follows obviously from Eq.~\ref{Sigma2_general}. Perhaps less obviously, phonon peaks representing multiphonon processes also arise, at energies $E_n+j\omega_E$ for integers $j\geq2$. These peaks occur because the factors $N(\omega\pm\omega_E)$ in the self-energy pick up not only the LLs, but also the phonon peaks in the preceding iteration. So, for example, $\Sigma_2^{(2)}(\omega)\propto N^{(1)}(\omega-\omega_E)$ will have a phonon peak at $2\omega_E$, corresponding to the phonon peak in $N^{(1)}$ at $\omega_E$. And each further iteration produces a new set of phonon peaks shifted further away from the Fermi energy by $\omega_E$. These new phonon peaks can cause further splitting of LLs. Moreover, the phonon peaks that were already present in the first iteration (shown in red) are shifted down in energy, causing split LLs at lower energies. Together, these effects of iteration cause the features in the DOS to become progressively more indistinct at energies $\gtrsim E_1+\omega_E$.

However, at energies below this, which are more relevant to current experiments, the only effect of iteration is a slight increase in $\lambda$; the qualitative features are wholly unaffected. And for smaller values of $A$, the effects are further reduced. For example, for the value $A=100$meV used in our fit to experimental data\cite{Pound:11}, iterating to convergence alters the parameters obtained from the fit by less than 1\%. Given this fact, in the following sections we will continue to use the Einstein model and only one iteration, since that affords the cleanest description of features in the DOS. But one should keep in mind two essential modifications that would occur in a more complete treatment: (i) near phonon peaks, a broadened phonon spectrum will smear features in the DOS; and (ii) in a converged solution, phonon peaks are displaced by $\omega_E$ relative to the \emph{renormalized} Landau levels rather than the bare ones.

\section{Density of states for finite chemical potential}\label{with chemical potential}
\subsection{Jumping phonon peaks}\label{jumping}

\begin{figure}[tb]
\begin{center}
\includegraphics[width=\columnwidth]{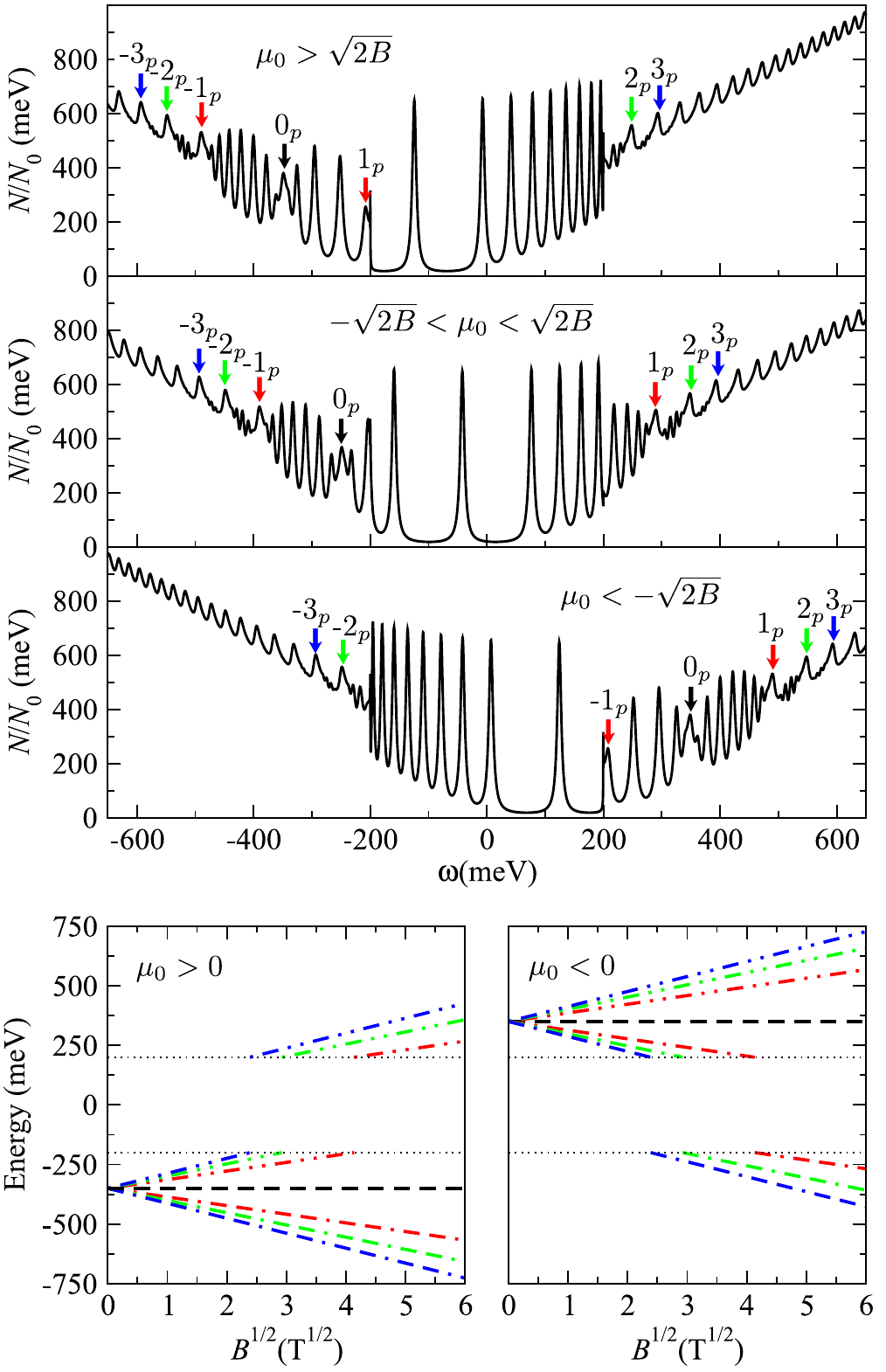}
\end{center}
\caption{(Color online) Top: Density of states as a function of energy for three finite values of $\mu$: 150meV~(top), 50meV~(middle), and -150meV~(bottom), chosen to lie in the intervals  $\mu_0>\sqrt{2B}$, $-\sqrt{2B}<\mu_0<\sqrt{2B}$, and $\mu_0<-\sqrt{2B}$, respectively. Arrows point to the locations of the phonon peaks. In all cases,  $B=15.2$T, $\omega_E=200$meV, $A=250$meV, and $\Gamma=5$meV. Bottom: the locations of the phonon peaks for $\mu_0=150$meV (left) and $\mu_0=-150$meV (right). The colors of the curves match those of the arrows in the top set of plots. Dotted lines indicate $\pm\omega_E$.\label{Fig:mu}}
\end{figure}

\begin{figure}[tb]
\begin{center}
\includegraphics[width=\columnwidth]{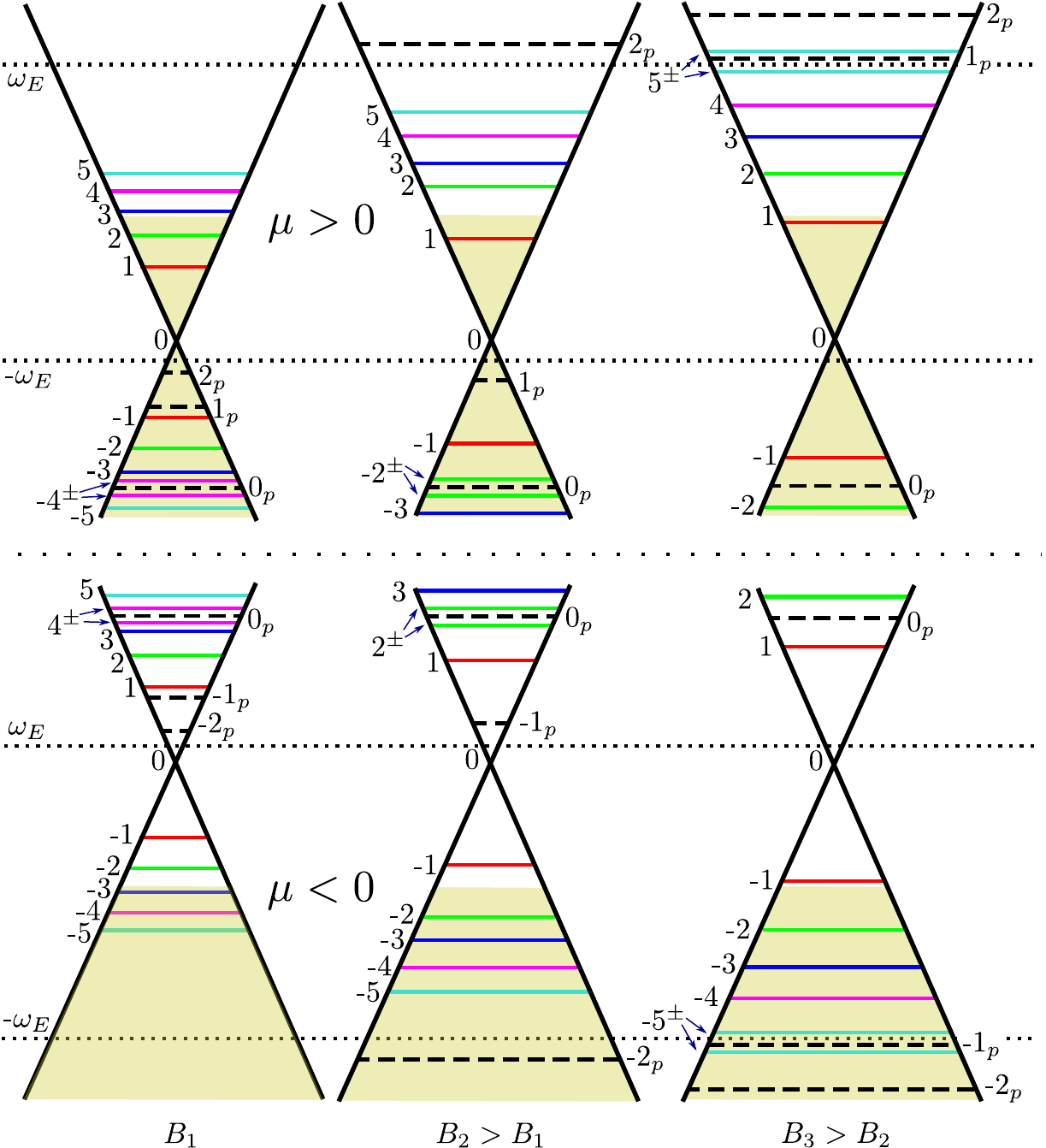}
\end{center}
\caption{(Color online) The arrangement of energy levels on the Dirac cone in the cases $\mu>0$ (top) and $\mu<0$ (bottom) for three increasing values of magnetic field. The shaded region indicates filling up to the Fermi energy. Landau levels are indicated by solid lines labeled with integers $n$ or $n^\pm$, and phonon peaks by dashed lines labeled with integers $m_p$ or $m_p^\pm$.\label{Fig:cones_mu}}
\end{figure}

When the chemical potential is finite, many new features appear in the DOS. First among these is a jump in the location of the phonon peaks as $B$ is varied. Referring back to Table~\ref{peaks}, we see that whenever a phonon peak at $P_m=M_m-\mu_0\pm\omega_E$ would fall in the window $(-\omega_E,\omega_E)$, its position is altered to $P_m=M_m-\mu_0\mp\omega_E$, causing it to jump from one side of the window to the other. This discontinuity in the positions of peaks is seen in the bottom frames of Fig.~\ref{Fig:mu}, which show plots of $P_m$ as a function of $\sqrt{B}$ for the two cases $\mu_0>0$ and $\mu_0<0$. Here we see that as $B$ is varied, the phonon peaks jump from one side of $\pm\omega_E$ (dotted black horizontal lines) to the other side of $\mp\omega_E$. More precisely, in the left plot, when $B>\mu_0^2/2$ none of the curves have reached $\pm\omega_E$, and they are given by $-\mu_0-\omega_E$ (black), $M_{\pm1}-\mu_0\pm\omega_E$ (red), $M_{\pm2}-\mu_0\pm\omega_E$ (green), and $M_{\pm3}-\mu_0\pm\omega_E$ (blue), where upper signs correspond to upper curves and lower to lower. When $B$ reaches $\frac{\mu_0^2}{2n}$, the $n$th energy above $\omega_E$, given by $M_n-\mu_0+\omega_E$, reaches $\omega_E$ and then jumps discontinuously to below $-\omega_E$, continuing on as $M_n-\mu_0-\omega_E$ at lower $B$. The energies in the right plot differ from those in the left only by a change of sign.

This phenomenon allows us to single out three cases, shown in the top three frames of Fig.~\ref{Fig:mu}: $\mu_0>\sqrt{2B}$, $-\sqrt{2B}<\mu_0<\sqrt{2B}$, and $\mu_0<-\sqrt{2B}$. In the first case, all phonon peaks with $m_p<0$ and at least one phonon peak with $m_p>0$ occur to the left of the window $(-\omega_E,\omega_E)$, while the remaining phonon peaks lie to the right; for example, in the top frame of the figure, $1_p$ lies to the left. In the second case, all phonon peaks with $m_p<0$ lie to the left and all with $m_p>0$ lie to the right. In the last case, at least one phonon peak with $m_p<0$ lies to the right; for example, $-1_p$ in the third frame. The position of the zeroth level relative to the window depends only on the sign of $\mu_0$: it lies to the left for $\mu_0>0$ and to the right for $\mu_0<0$. As seen in the preceding section, for $\mu_0=0$, the peak is split into two, with one at $-\omega_E$ and one at $\omega_E$. In all cases, the logarithmic peaks lie at $\pm\omega_E$.

Figure~\ref{Fig:cones_mu} shows this jumping behavior schematically on the Dirac cone. Just as in the corresponding schematic for $\mu=0$, shown in Fig.~\ref{Fig:cones}, solid colored lines labeled with integers $n$ are Landau levels and dashed black lines labeled with integers $m_p$ are phonon peaks. The cones are now shifted by $-\mu$ relative to the Fermi energy (the filled region below the Fermi energy is shaded), so in the upper set of cones, for $\mu>0$, the cones are shifted downward, and in the lower set, for $\mu<0$, the cones are shifted upward. As in the earlier schematic, both the LLs and the phonon peaks move outward from the Dirac point as $B$ increases, and the LLs split when passing through a phonon peak. However, now as we increase $B$, the locations of the phonon peaks vary discontinuously. On the upper set of cones, we see first $2_p$ and then $1_p$ jump from below $-\omega_E$ to above $\omega_E$. The lower set is the mirror of the upper, so as we increase $B$, we see first $-2_p$ and then $-1_p$ jump from above $\omega_E$ to below $-\omega_E$.

\subsection{Large chemical potential and split zeroth level}

\begin{figure}[tb]
\begin{center}
\includegraphics[width=\columnwidth]{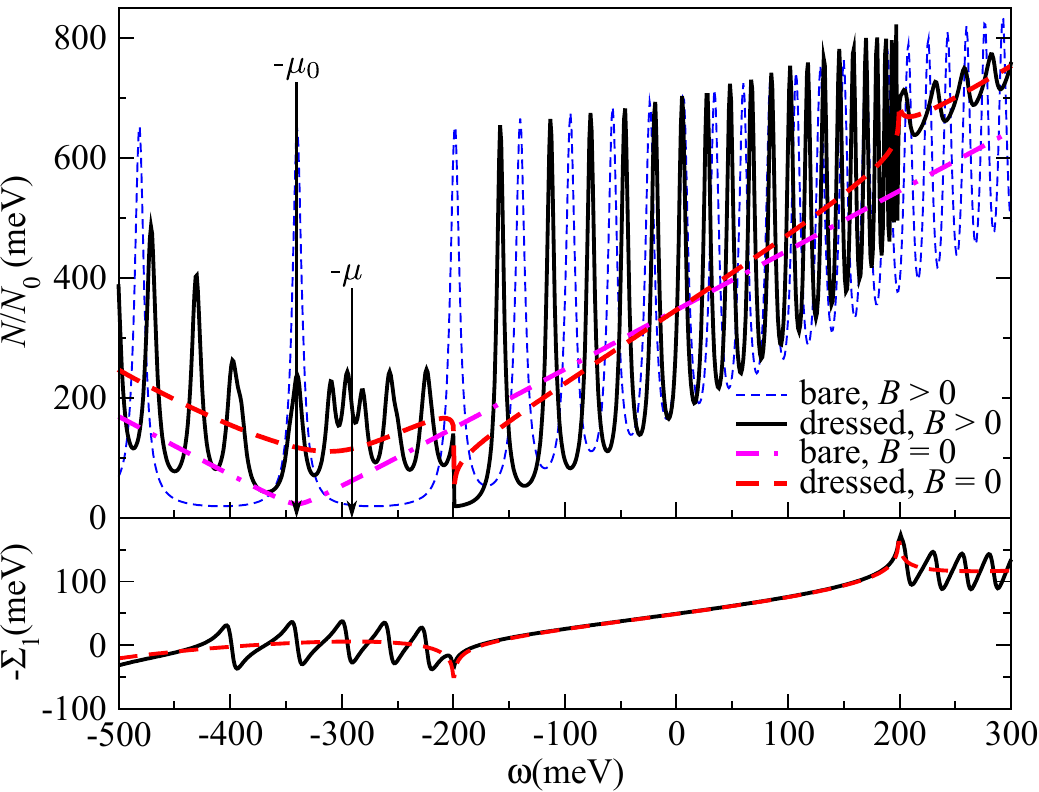}
\end{center}
\caption{(Color online) Bare (thin dashed blue) and renormalized (solid black) density of states and the real part of the self-energy in the case of $\mu>\omega_E$. Arrows point to the (negative of the) bare and renormalized chemical potentials $\mu_0=340$meV and $\mu=291$meV, respectively. The remaining parameters are $\omega_E=200$meV, $A=250$meV, $B=15.2$T, and $\Gamma=5$meV. We also display the bare (dot-dashed magenta) and renormalized (thick dashed red) curves in the $B=0$ case.\label{Fig:mu_large}}
\end{figure}

\begin{figure}[tb]
\begin{center}
\includegraphics[width=\columnwidth]{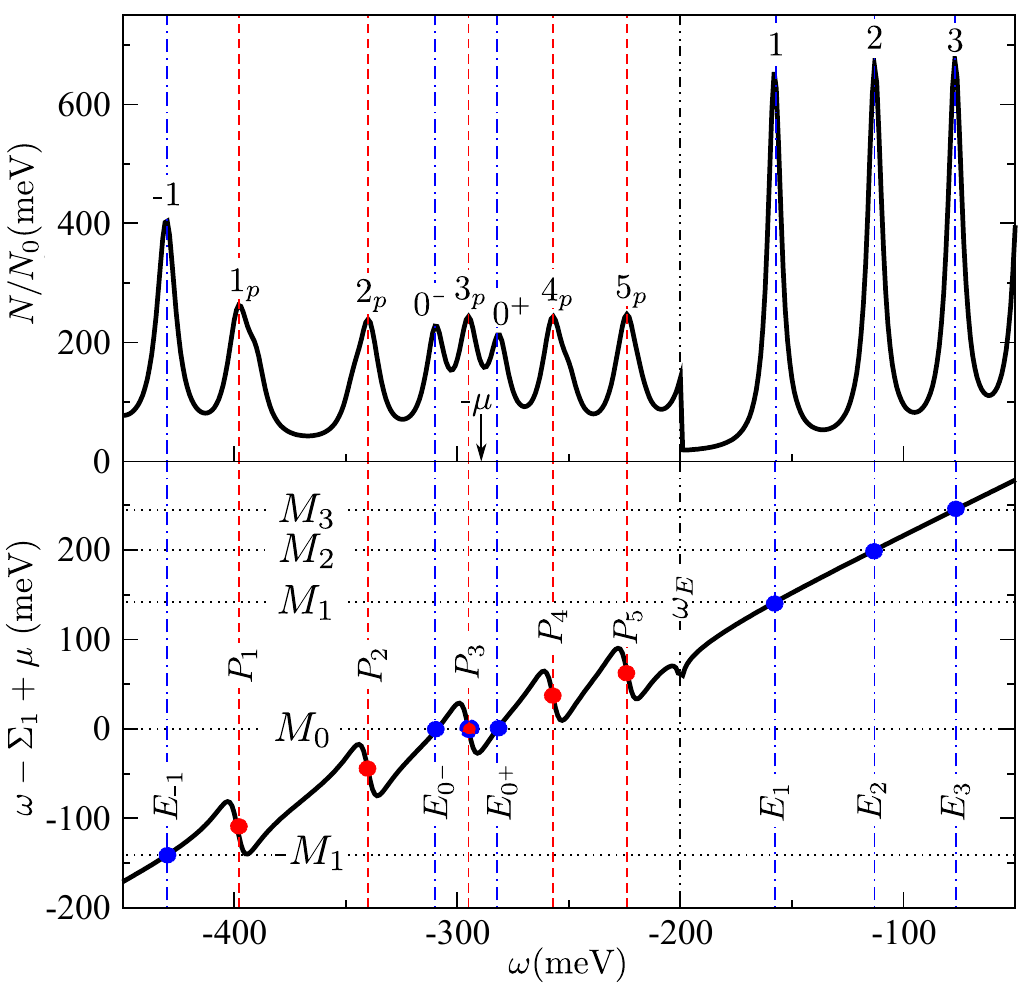}
\end{center}
\caption{(Color online) A small portion of Fig.~\ref{Fig:mu_large} focusing on the region around $-\mu$. Here we see that the three peaks near $-\mu$ consist of a phonon peak and the split $n=0$ Landau level. None of the three lies at $-\mu$.\label{Fig:mu_peak_locations}}
\end{figure}

\begin{figure}[tb]
\begin{center}
\includegraphics[scale=0.8]{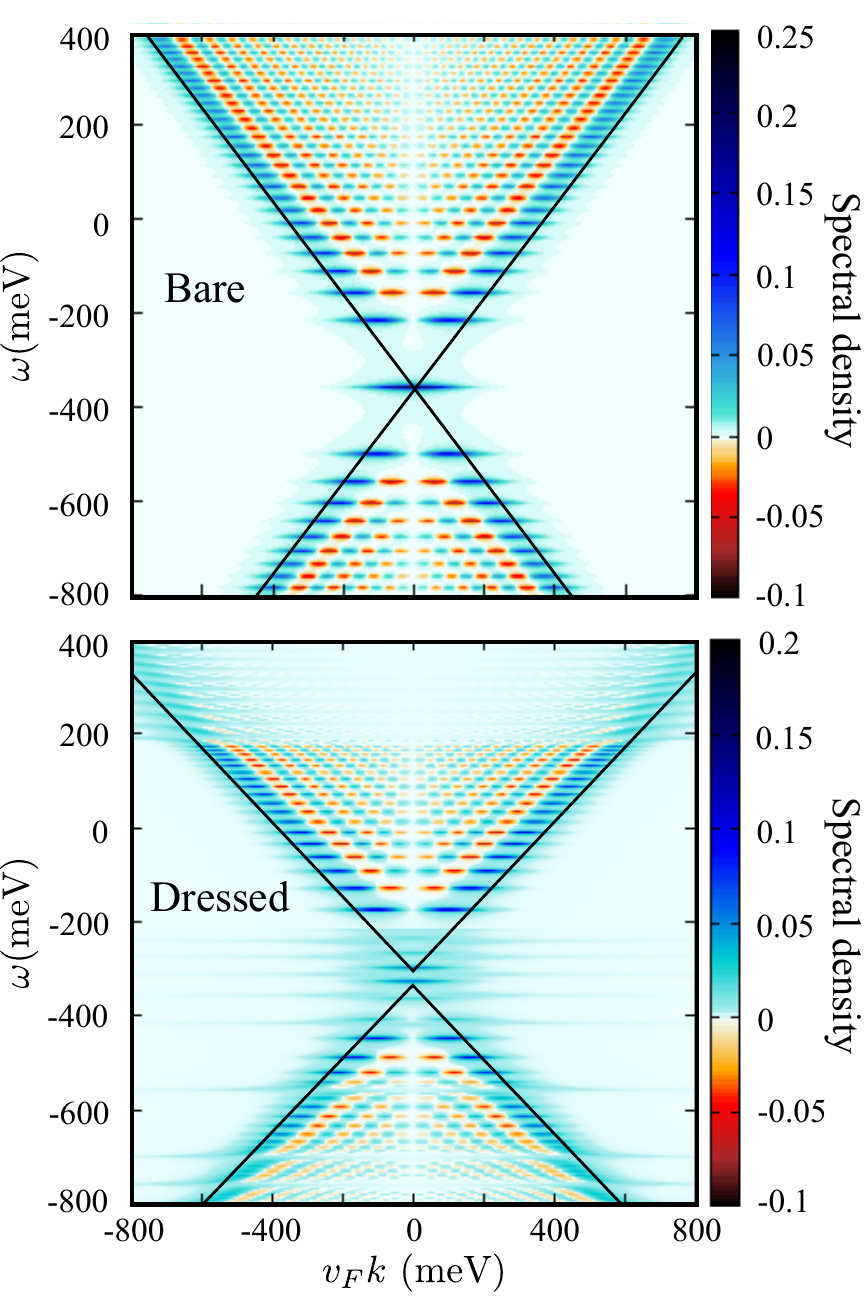}
\end{center}
\caption{(Color online) Color maps of the bare (top) and dressed (bottom) spectral density ${\rm tr}[\gamma^0A(\vec k,\omega)]$ in the $(k,\omega)$ plane for $\mu_0=-340$meV, $\omega_E=200$meV, $A=250$meV, $B=15.2$T, and $\Gamma=3$meV. The plane is extended to negative values of $k$ in order to show the cone-like structure. \label{Fig:spectral_density}}
\end{figure}

When $|\mu|>\omega_E$, the $n=0$ Landau level lies outside the window $(-\omega_E,\omega_E)$, introducing several new features. This scenario is shown in Fig.~\ref{Fig:mu_large}, where we show both the bare and dressed DOS. We also show the $B=0$ curves for comparison. For $B=0$, the DOS at the Dirac point (at $-\mu_0$ in the bare case and slightly to the left of $-\mu$ in the dressed case) is pronouncedly lifted by renormalization. In the finite-$B$ case, this occurs as well, and we see here that the valleys around $-\mu$ are no longer global minima, as they were in the bare case. But additional, even more marked changes occur in the finite-$B$ case: there is no single peak near $-\mu$ that indicates the Dirac point. Recall that since $\Sigma_1$ is effectively unchanged by the presence of a magnetic field in the window $(-\omega_E,\omega_E)$, the chemical potential $\mu=\mu_0+\Sigma_1(0)$ is also unaffected by the field, so $-\mu_0$ and $-\mu$ refer to both the $B=0$ and $B>0$ cases.

We examine this situation more closely in Fig.~\ref{Fig:mu_peak_locations}, which shows the region around $-\mu$ in more detail, labeling all the peaks in the region. We see that the three-peak structure near $-\mu$ consists of a phonon peak, $3_p$, and a split zeroth LL. Hence, when $\mu$ lies outside the window $(-\omega_E,\omega_E)$, the zeroth LL loses much of its significance: it may be split just as any other LL is when sufficiently close to a phonon peak, and it is not pinned to the Dirac point. (Also note the distinct character of the peak at $-\omega_E$, arising from the logarithmic terms in $\Sigma_1$, which for these particular parameters is clearly visible because none of the phonon peaks' energies $P_m$ lie near $-\omega_E$.)

Since there is no single zeroth LL, we have no clear way of identifying a Dirac point here. In the $B=0$ case, we can define the Dirac point even when we include interactions. The bare band energies $\epsilon_k=\pm v_Fk-\mu_0$ become the dressed energies $E_k$, given by $E_k-\Sigma_1(E_k)=\pm v_F k-\mu$. We can then define the Dirac point as the point of $k=0$ on these curves, such that its energy $\omega_d$ satisfies $\omega_d-\Sigma_1(\omega_d)=-\mu$. Although the dispersion relation is only approximately linear in this case, it is sufficiently close to linearity for the Dirac point to remain meaningful. In the finite-$B$ case, however, the linear bands vanish, replaced by discrete LLs. Even when neglecting interactions, this seems to make the definition of the Dirac point ambiguous. We illustrate this situation in Fig.~\ref{Fig:spectral_density}, which shows color maps of the spectral density as a function of $k$ and $\omega$; the black lines are intended only to guide the eye. The spectral density here is given by\cite{Sharapov:04}
\begin{align}
{\rm tr}[\gamma^0A(\vec{k},\omega)] &= 2e^{-v_F^2k^2/B}\sum_{n=0}^\infty(-1)^n\nonumber\\
&\quad\times\left[L_n(2v_F^2k^2/B)-L_{n-1}(2v_F^2k^2/B)\right]\nonumber\\
&\quad\times\frac{1}{\pi}\left[\frac{\Gamma-\Sigma_2}{(\omega-\Sigma_1+\mu-M_n)^2+(\Gamma-\Sigma_2)^2}\right.\nonumber\\
&\quad\left.+\frac{\Gamma-\Sigma_2}{(\omega-\Sigma_1+\mu+M_n)^2+(\Gamma-\Sigma_2)^2}\right],
\end{align}
where the spectral function $A(k,\omega)$ is a 4-by-4 block-diagonal matrix with one block for the $K$ point and one for the $K'$ point, $\gamma^0=\sigma_3\otimes\sigma_3$ is the Dirac matrix (where $\sigma_3$ is the Pauli matrix), $k=|\vec{k}|$, and $\hbar=1$. As we see in the upper frame of the figure, in the bare case the LLs are arranged on two fairly definite cones that intersect at the $n=0$ level. So although there are no longer linear bands, we can still define a Dirac point based on this conic arrangement of levels, and we can identify the $n=0$ level with the Dirac point. But in the dressed case, shown in the lower frame, the conic arrangement is radically distorted, particularly by the splitting of the $n=0$ level. This leaves considerable ambiguity in defining the cones or a Dirac point. However, we can see from the figure that two intersecting lines alone cannot be made to pass through the peaks in the spectral function. In that sense, we can say that the splitting of the zeroth LL corresponds to a splitting of the Dirac point into two.

Although this picture of the spectral density would be altered if electron-electron interactions were taken into account as well, we expect it to remain qualitatively valid. In the figure we see each energy level broken into a series of islands of strong spectral weight, due to the Laguerre polynomials in the spectral function. (These islands arise from the multiplication of the two sublattice components in the wavefunction in a magnetic field, the $K$-point piece of which reads\cite{Peres:06} $\Psi\sim\sum_n(\varphi_{|n|},\varphi_{|n|+1})$ in the $(A,B)$ sublattice basis, where $\varphi_n$ is the harmonic oscillator eigenfunction.) Below $-\omega_E$, we see that the islands are smeared at the energies of phonon peaks; above $\omega_E$, the islands become indistinguishable because their spacing is on the same scale as the phonon peaks. If electron-electron interactions were accounted for, the particle-hole excitations would also be localized into islands,\cite{Roldan:10} with a spacing in $\omega$ equal to the difference between two LLs and a spacing in $k$ determined by the wavefunction of the lower level, which would cause smearing at those locations. But unlike in the zero-field case, the excitations are gapped out at the Dirac point, so the splitting of the $n=0$ level due to electron-phonon coupling could be distinguished.

\begin{figure}[tb]
\begin{center}
\includegraphics[width=\columnwidth]{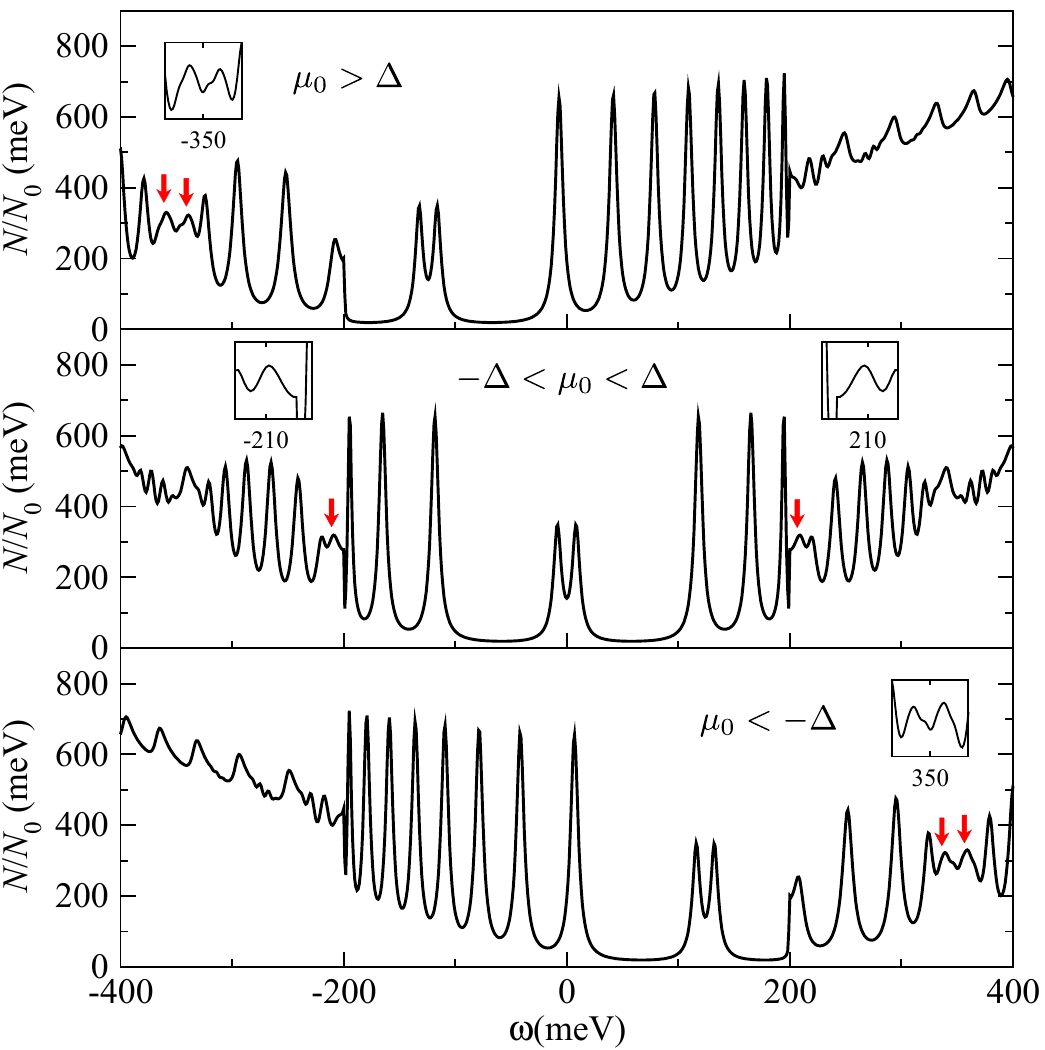}
\end{center}
\caption{(Color online) Density of states with a finite energy gap $\Delta=10$meV for three values of $\mu_0$: 150meV~(top), 0meV~(middle), and -150meV~(bottom), chosen to lie in the intervals $\mu_0>\Delta$, $-\Delta<\mu_0<\Delta$, and $\mu_0<-\Delta$, respectively. Red arrows point to the locations of the phonon peaks around $\pm\omega_E$, shown in the insets. For $-\Delta<\mu_0<\Delta$, these peaks lie at $\pm(\omega_E+\Delta)-\mu_0$; for $\mu_0>\Delta$, at $-\omega_E\pm\Delta-\mu_0$; for $\mu_0<-\Delta$, at $\omega_E\pm\Delta-\mu_0$. In all cases, $\omega_E=200$meV, $A=250$meV, $B=15.2$T, and $\Gamma=5$meV.\label{Fig:Delta}}
\end{figure}

\begin{figure}[tb]
\begin{center}
\includegraphics[width=\columnwidth]{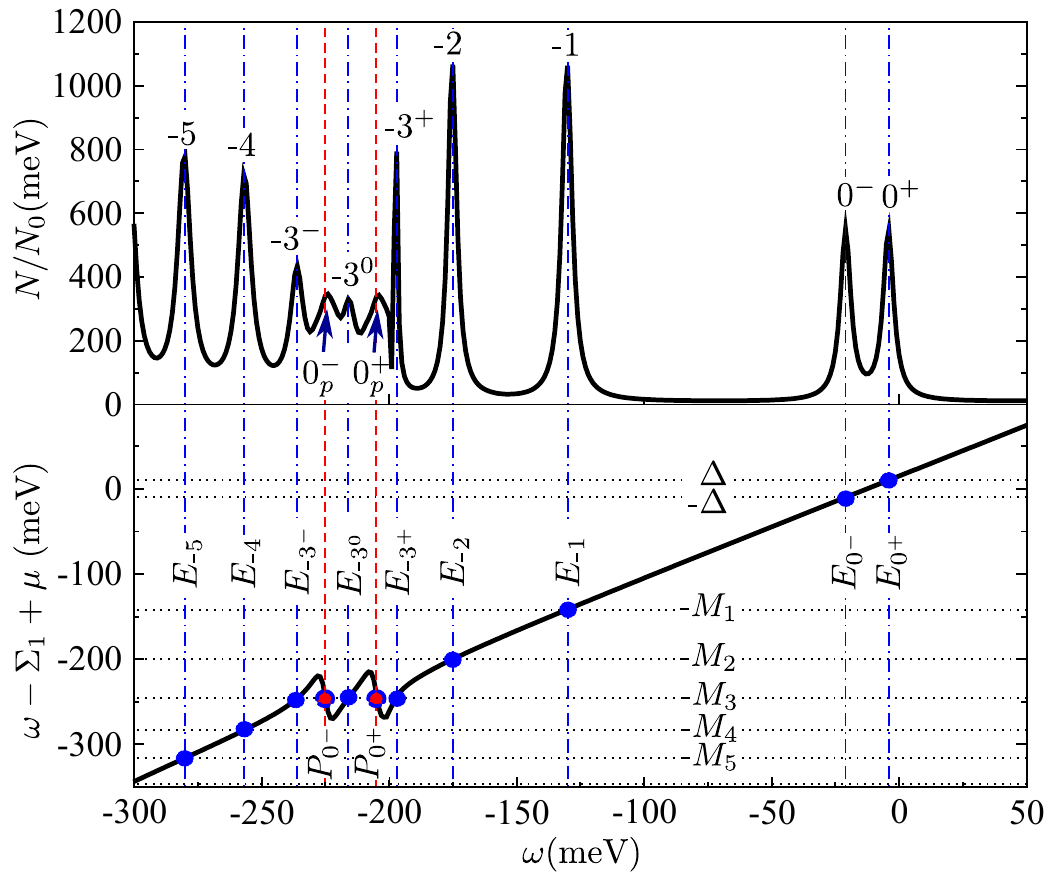}
\end{center}
\caption{(Color online) A density of states curve displaying a Landau level ($n=-3$) split in three by a phonon peak that is itself split in two by an energy gap. Here $\mu_0=15$meV, $\Delta=10$meV, $A=250$meV, $\omega_E=200$meV, $\Gamma=3$meV, and $B=15.2$T.\label{Fig:Delta_splitting}}
\end{figure}

\begin{figure}[htb]
\begin{center}
\includegraphics[width=\columnwidth]{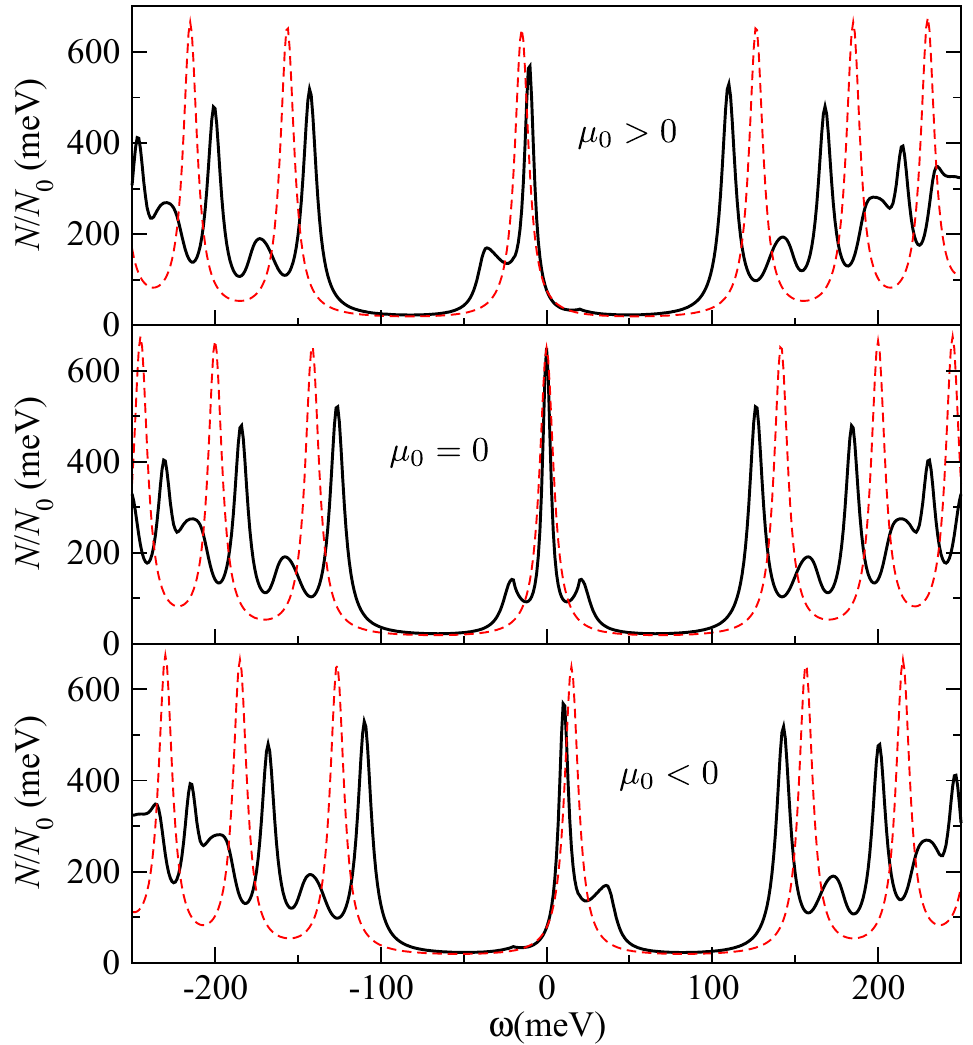}
\end{center}
\caption{(Color online) The bare density of states (dashed red curves) and the density of states renormalized by an acoustic phonon (solid black curves), for three values of $\mu_0$: 15meV~(top), 0meV~(middle), and -15meV(bottom). In all cases, $B=15.2$T, $\Gamma=5$meV, and the parameters of the Debye distribution are $\omega_D=20$meV and $A'=0.03125$meV$^{-2}$.\label{Fig:Debye}}
\end{figure}

\section{Other effects creating new peaks}\label{other}
In this section, we analyse two more ways in which phonons can induce new peaks in the DOS. First, we discuss how coupling to a phonon causes new structures to arise from the splitting of the zeroth LL by an energy gap. Next, we discuss the effects of an acoustic phonon modeled by a Debye distribution.

\subsection{Finite energy gap}\label{gap}
Recent experimental data has suggested the existence of an energy gap $\Delta$ at the Dirac point in graphene.\cite{Li-G:09} The cause of this gap is unknown, but may be related to sublattice asymmetry.\cite{Khveshchenko:01,Gusynin:06a} As we show in this section, the signature of a gap appears not only in splitting the $n=0$ Landau level, but in the images $0_p^\pm$ of that split level in the phonon peaks at $\pm\omega_E$. From Table~\ref{peaks}, we see that the locations of the $0^\pm_p$ levels depend on the relative size of $\Delta$ and $\mu_0$. The three possible cases are shown in Fig.~\ref{Fig:Delta}, with the locations of the peaks indicated by red arrows. If $\mu_0$ lies between $-\Delta$ and $+\Delta$, as shown in the middle frame of the figure, then the image of the bare level $n=0^+$ at $\omega=\Delta-\mu_0$ appears to the right of $\omega_E$, as the $0_p^+$ level at $\omega=\Delta-\mu_0+\omega_E$, while the image of the bare level $n=0^-$ at $\omega=-\Delta-\mu_0$ appears to the left of $-\omega_E$, as the $0_p^-$ level at $\omega=-\Delta-\mu_0-\omega_E$. If $\mu_0$ lies outside this range, then both the $0^+_p$ and $0^-_p$ levels appear on the same side of the window $(-\omega_E,\omega_E)$: to the right of $\omega_E$, at $\pm\Delta-\mu_0+\omega_E$, if $\mu_0>\Delta$; or to the left of $-\omega_E$, at $\pm\Delta-\mu_0-\omega_E$, if $\mu_0<-\Delta$. These two cases are shown in the top and bottom frames of Fig.~\ref{Fig:Delta}, respectively.

If the $0_p^\pm$ peaks both lie on the same side of the window $(-\omega_E,\omega_E)$, then any sufficiently close Landau level will be split into three peaks rather than two. We illustrate this in Fig.~\ref{Fig:Delta_splitting}, where the LL $n=-3$ is split in three. Because the bare $n=0^\pm$ levels have equal magnitude, their images $0_p^\pm$ do as well, creating the two neighbouring oscillations of equal magnitude seen in $\omega-\Sigma_1+\mu$. For sufficiently small $\Delta$, this means that the splitting of a LL into three is generic, since a line of constant $M_n$ will have to intersect both oscillations if it intersects either.

\subsection{Acoustic phonon}

Recent STS experiments have also observed shoulders and distinct secondary peaks on the Landau levels.\cite{Miller:09} While some or all of these peaks may be caused by the STM tip deforming the graphene sheet,\cite{Kubista:11} they could potentially be caused by coupling to an acoustic phonon or some other low-energy boson. To consider this case, rather than the Einstein model, we adopt a Debye distribution for the electron-phonon spectral density,
\begin{equation}
\alpha^2 F(\nu) = A'\nu^2\theta(\omega_D-\nu).
\end{equation}
Figure~\ref{Fig:Debye} shows the DOS for three cases: $\mu_0>0$, $\mu_0=0$, and $\mu_0<0$. When the chemical potential vanishes, the acoustic phonon manifests as a hump on either side of the zeroth LL, along with humps at $M_n\pm\omega_D$. When the chemical potential is finite, only one hump appears near the zeroth LL, to the left for $\mu_0>0$ and to the right for $\mu_0<0$, in the same manner that $P_0$ appeared to the right or left in the case of the Einstein model at finite $\mu$. The other humps are unaffected, only shifted by $-\mu_0$ to $M_n-\mu_0\pm\omega_D$. 

From this, we see that coupling to a Debye distribution has qualitatively the same effect as coupling to a single Einstein mode, introducing a new set of peaks that is shifted relative to the Landau levels by the characteristic phonon frequency (in this case, the Debye frequency). However, since the Debye distribution is finite at all nonzero frequencies below $\omega_D$, in the Debye model the interval $(-\omega_D,\omega_D)$ is `softer' than the analogous interval $(-\omega_E,\omega_E)$ in the Einstein model. By that we mean that phonon peaks can appear within the window, but with greatly reduced magnitudes, and LLs within the window will be more affected by the phonon. But the Debye frequency is likely to be small compared to the scale of the LLs in experiment, so the window will contain few levels in any case.

\section{Conclusion}\label{conclusion}
In this paper, we have shown how in the presence of a magnetic field $B$, electron-phonon coupling induces a new set of peaks in the electronic density of states in graphene, over and above the Landau levels in the bare DOS; furthermore, we have shown how these new peaks affect the locations of the renormalized LLs themselves. We can understand the location of the new peaks, which we have called phonon peaks, by noting that the electron-phonon self-energy $\Sigma$ is associated with an electron (or hole) scattering between states of energy $\omega$ and $\omega-\omega_E$ (or $\omega+\omega_E$), where $\omega_E$ is the phonon energy. This means that the self-energy at $\omega$ contains a direct imprint of the DOS at $\omega\pm\omega_E$, leading to peaks in $\Sigma$ at energies displaced by $\pm\omega_E$ from the LLs. Renormalization then introduces these phonon peaks into the dressed DOS. Since this scattering process is possible only for $|\omega|>\omega_E$, the phonon peaks always lie outside the window $(-\omega_E,\omega_E)$. The presence of these new peaks alters the behavior of the LLs as well. When a LL lies far from a phonon peak, renormalization simply scales its energy by a factor of $1/(1+\lambda^{\rm eff})$ (equivalent to a  renormalization of the Fermi velocity), where $\lambda^{\rm eff}$ is the effective mass renormalization parameter. However, the nearer a LL lies to a phonon peak, the more it deviates from this behavior; and if it is sufficiently near, it splits in two about the phonon peak. Because the phonon peaks lie outside the window $(-\omega_E,\omega_E)$, LLs within that window are left almost entirely intact, simply shifted by the renormalization of the Fermi velocity.

This picture is made more complex in the case of a finite chemical potential, when the Landau levels and phonon peaks are no longer centered on the Fermi energy. Because no phonon peaks can occur in the window $(-\omega_E,\omega_E)$, varying $B$ or $\mu$ continuously causes the phonon peaks to jump discontinously from one side of the window to the other. And if the chemical potential is greater than the phonon frequency, then the zeroth LL lies outside the window, falling amongst the phonon peaks. This means that it can be split in two, losing its signifcance as an identifier of the Dirac point.

We have also discussed how new peaks arise in the presence of an energy gap at the Dirac point or a coupling to an acoustic phonon. In the first case, the energy gap splits the bare zeroth LL, leading to a split zeroth phonon peak $P_{0^\pm}$. Depending on the value of the chemical potential, the peaks $P_{0^\pm}$ can both lie to one side of the window $(-\omega_E,\omega_E)$, or one can lie to either side of it. If both lie on the same side, they can split a nearby LL in three. In the second case, coupling to an acoustic phonon modeled by the Debye distribution has the same qualitative effect as coupling to an Einstein mode, introducing phonon peaks at $M_n\pm\omega_D$. In the case of charge neutrality, this leads to two phonon peaks centered about the zeroth LL; for finite chemical potential, there is only one zeroth phonon peak, lying to the right or left of the zeroth LL.

Together, these analyses reveal a rich electronic structure that emerges from an interplay between Landau level quantization and electron-phonon coupling in graphene. Because the features in this structure are tied to a fixed frequency that doesn't depend on magnetic field, they should be observable in experiment even in the presence of complicating factors such as electron-electron interactions, finite size effects, and substrate effects. Indeed, some of the features have already been observed in STS data;\cite{Pound:11} others, such as the hopping of phonon peaks at finite $\mu$, should be readily observable, while precise observations of some, such as the behavior of split Landau levels, may require higher quality samples. And future studies of different quantities, such as optical conductivity, should yield similarly rich results.

\bibliography{DOS_long}
\end{document}